\documentclass[pdflatex,sn-mathphys-num]{sn-jnl}

\usepackage{graphicx}%
\usepackage{multirow}%
\usepackage{amsmath,amssymb,amsfonts}%
\usepackage{amsthm}%
\usepackage{mathrsfs}%
\usepackage[title]{appendix}%
\usepackage{xcolor}%
\usepackage{textcomp}%
\usepackage{manyfoot}%
\usepackage{booktabs}%
\usepackage{algorithm}%
\usepackage{algorithmicx}%
\usepackage{algpseudocode}%
\usepackage{listings}%
\usepackage{overpic}%
\usepackage{bm}%

\theoremstyle{thmstyleone}%
\theoremstyle{thmstyletwo}%

\theoremstyle{thmstylethree}%

\begin{document}

\title[Article Title]{Effect of flow kinematics on extensional viscosity of dilute polymer solutions}


\author*[1]{\fnm{Yusuke} \sur{Koide}}\email{koide.yusuke.k1@f.mail.nagoya-u.ac.jp}

\author[1]{\fnm{Takato} \sur{Ishida}}

\author[1]{\fnm{Takashi} \sur{Uneyama}}
\author[1]{\fnm{Yuichi} \sur{Masubuchi}}

\affil[1]{\orgdiv{Department of Materials Physics}, \orgname{Nagoya University}, \orgaddress{\city{Nagoya}, \state{Aichi}, \country{Japan}}}




\abstract{
We investigate the effect of flow kinematics on the extensional viscosity of dilute polymer solutions by conducting dissipative particle dynamics simulations under uniaxial, planar, and biaxial extensional flows.
At high extension rates, dilute polymer solutions exhibit strain hardening under these flows, while the quantitative behavior depends on the flow type.
To elucidate the physical origin of this flow-kinematics dependence, we relate the extensional viscosity to polymer conformation using an analytical expression derived from a single-chain model.
The resulting relation allows us to separate the contribution of flow-induced polymer conformational changes and the purely kinematic contribution determined by the structure of the velocity gradient tensor.
When polymers remain almost unperturbed by extensional flows, differences in the extensional viscosity are governed primarily by the purely kinematic effects.
In contrast, as polymers are stretched, the gyration radius in the extensional direction becomes the dominant factor, and differences in the stretching degree in this direction lead to corresponding variations in the extensional viscosity. 
}


\maketitle

\section{Introduction}\label{Introduction}
Dilute polymer solutions have attracted considerable attention in fluid mechanics because even small amounts of polymer additives can qualitatively alter flow behavior.
Since Toms's experiments in the 1940s~\cite{Toms}, it has been known that long, flexible polymers can drastically suppress turbulence, thereby contributing to drag reduction in wall-bounded flows~\cite{White2008-gb,Xi2019-nc}.
In contrast, under laminar-flow conditions, polymers can induce elastic turbulence, enhancing mixing efficiency in microfluidic devices~\cite{Datta2022-vl,Sasmal2025-pb}.
To reveal the physical mechanism underlying these intriguing phenomena, a number of researchers have studied flow modulation by polymers through laboratory experiments and numerical simulations based on constitutive equations, such as the Oldroyd-B model and the finitely extensible nonlinear elastic model with the Peterlin approximation~(FENE-P).

In this context, a fundamental understanding of the rheology of dilute polymer solutions is indispensable because it serves as a basis for interpreting complex flow data and assessing the validity of constitutive equations.
In realistic situations~(e.g., turbulent flows), polymers undergo various flow types, such as shear and extensional flows, as well as their combinations.
Moreover, the polymer response is highly sensitive to the nature of the flow kinematics.
For example, polymers exhibit significant stretching under extensional flow~\cite{Perkins1997-dm}, in contrast to their behavior under shear flow~\cite{Smith1999-rk}.
Thus, it is crucial to understand the rheological properties of dilute polymer solutions in different flow types.
In particular, since the shear viscosity of dilute polymer solutions is typically independent of the shear rate or exhibits only weak shear thinning, the extensional rheology has been a major focus. 

However, due to the low viscosity of dilute polymer solutions, systematic investigations under extensional flows with different flow kinematics remain scarce~\cite{Jones1987-gx,Haward2023b}, compared with polymer melts~\cite{Khan1987-br,Takahashi1993-mb,Nishioka2000-eh,Wagner2001-by,Sugimoto2001-ji,Hachmann2003-di,Stadler2007-gq}.
Recently, extensional rheometry techniques for low-viscosity fluids, such as the capillary breakup extensional rheometer~\cite{McKinley2000-ky,Anna2001-bn} and cross-slot extensional rheometry~\cite{Haward2016-lh}, have been developed, enabling evaluation of the extensional rheology of dilute polymer solutions.
Notably, Haward et al.~\cite{Haward2023a} developed a six-arm cross-slot device optimized for generating uniaxial and biaxial extensional flows. 
Together with the two-dimensional cross-slot device, they systematically investigated the extensional viscosity of dilute polymer solutions under uniaxial~($\eta_\mathrm{E}(\dot{\epsilon})$), planar~($\eta_\mathrm{P}(\dot{\epsilon}_\mathrm{P})$), and biaxial extensional flows~($\eta_\mathrm{B}(\dot{\epsilon}_\mathrm{B})$)~\cite{Haward2023b}.
They reported that at low concentrations of polymers, $\eta_\mathrm{E}(\dot{\epsilon})$, $\eta_\mathrm{P}(\dot{\epsilon}_\mathrm{P})$, and $\eta_\mathrm{B}(\dot{\epsilon}_\mathrm{B})$ are consistent with the predictions of the FENE-P model, including the relation $\eta_\mathrm{E}(\dot{\epsilon})=\eta_\mathrm{P}(\dot{\epsilon}_\mathrm{P})=2\eta_\mathrm{B}(\dot{\epsilon}_\mathrm{B})$ at high extension rates.
While these approaches are promising, further work is needed to establish quantitative reliability, particularly with respect to nonidealities in the applied flow fields and the approximations used in the viscosity calculations.

Molecular simulations can complement experimental measurements and provide valuable insights into the relation between extensional rheology and polymer dynamics.
There have been many molecular simulation studies of dilute polymer solutions under uniaxial and planar extensional flows~\cite{van-den-Brule1993-mg,Doyle1997-tq,Herrchen1997-fb,Li2000-da,Larson2002-mw,Somasi2002-ot,Hsieh2003-oc,Larson2005-ef}.
In our previous work~\cite{koide2025relation}, we conducted dissipative particle dynamics~(DPD) simulations of dilute polymer solutions under uniaxial extensional flow and established a quantitative relation between extensional viscosity and polymer conformation.
However, a systematic comparison of the rheological response under different extensional flows, especially under biaxial extensional flows, remains unexplored for dilute polymer solutions, although the effect of biaxial deformation has been studied for polymer melts~\cite{Takeda2015-ub,Murashima2018-te,Murashima2021-wh}.
This is a serious issue because several previous studies reported that biaxial extensional flows play a crucial role in polymer stretching in turbulent flows~\cite{Zhou2003-gk,Terrapon2004-xb}.

In the present study, we aim to establish a comprehensive understanding of the effect of flow kinematics on dilute polymer solutions.
For this purpose, we conduct DPD simulations of dilute polymer solutions under uniaxial, planar, and biaxial extensional flows and compare the extensional viscosities among these flow types.
To explain the flow-kinematics dependence of the rheological response, we relate the extensional viscosity to polymer conformations.
Specifically, we extend our previous approach for uniaxial extensional flow~\cite{koide2025relation}, which is based on an analytical relation for a single-chain model, to planar and biaxial extensional flows.
The quantitative relations obtained here enable us to decouple the contributions of flow-induced polymer conformational changes from the purely kinematic effects determined by the structure of the velocity gradient tensor.
Consequently, we reveal how flow kinematics affects the extensional viscosity of dilute polymer solutions in terms of polymer conformation.

\section{Method}\label{Method}
We employ the DPD method to study dilute polymer solutions.
Following previous studies~\cite{Jiang2007-rf,koide2025relation}, polymers are modeled as flexible linear chains, each of which consists of $N_\mathrm{p}$ particles.
Adjacent particles along the chain are connected by a shifted FENE potential $U_\mathrm{B}$:
\begin{equation}
  U_\mathrm{B}(r) = -\frac{k_F}{2}(r_\mathrm{max}-r_\mathrm{eq})^2\ln \left\{1-\left (\frac{r-r_\mathrm{eq}}{r_\mathrm{max}-r_\mathrm{eq}}\right )^2\right\},
\end{equation}
where $k_F$ is the spring constant, $r_\mathrm{eq}$ is the equilibrium bond length, and $r_\mathrm{max}$ is the maximum bond length.
In the DPD method, solvent particles are considered explicitly.
These DPD particles, including polymer and solvent particles, interact via soft repulsive, dissipative, and random forces. 
The dissipative force, which depends on the relative velocity of interacting particles, is computed using the laboratory velocity.
Further details of the DPD model can be found in previous papers~\cite{Koide2022-bp,Koide2023-ao}.
In the following, all quantities are nondimensionalized using $k_BT$, $m$, and $r_c$, where $k_B$ is the Boltzmann constant, $T$ is the temperature of the system, $m$ is the mass of a DPD particle, and $r_c$ is the cutoff distance.

The parameters of the DPD simulations are set as follows:
the total number of DPD particles is $N=648\,000$; the number density of particles is $\rho=3$; the dissipative force coefficient is $\gamma=4.5$; the random force coefficient is $\sigma=3$; the number of particles per chain is $N_\mathrm{p}=50$; the volume fraction of polymer particles is $\phi=0.1$; the spring constant is $k_F=40$; the equilibrium bond length is $r_\mathrm{eq}=0.7$; the maximum bond length is $r_\mathrm{max}=2$; the repulsive force coefficients between different types of particles are $a_\mathrm{pp}=25$, $a_\mathrm{ps}=25$, and $a_\mathrm{ss}=25$, where p and s denote polymer and solvent particles, respectively.
We have confirmed that $\phi=0.1$ is within the dilute regime $\phi<\phi^*(=0.14)$, where the overlap volume fraction $\phi^*$ is computed as $\phi^* = {N_\mathrm{p}}/\{(4/3)\pi \rho R_{g}^3\}$ with $R_{g}$ being the polymer gyration radius evaluated at $\phi= 0.025$.

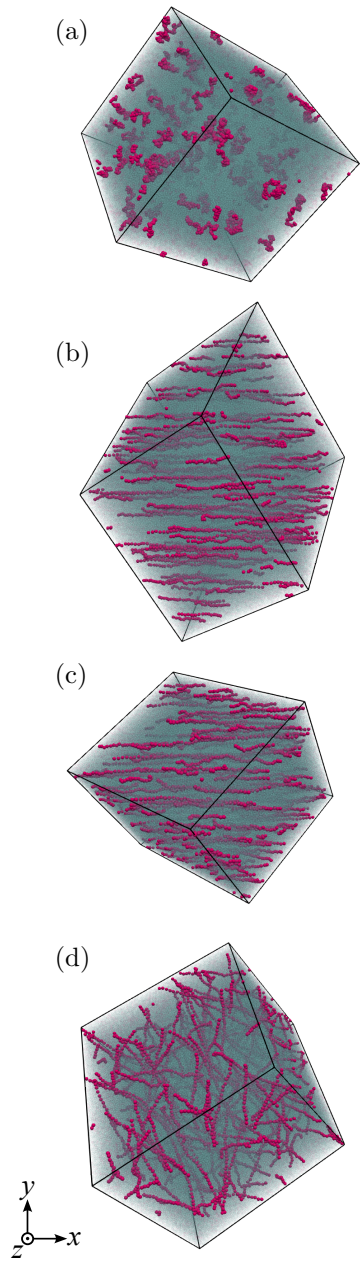
\begin{figure}
  \centering
  \begin{overpic}[width=0.4\linewidth]{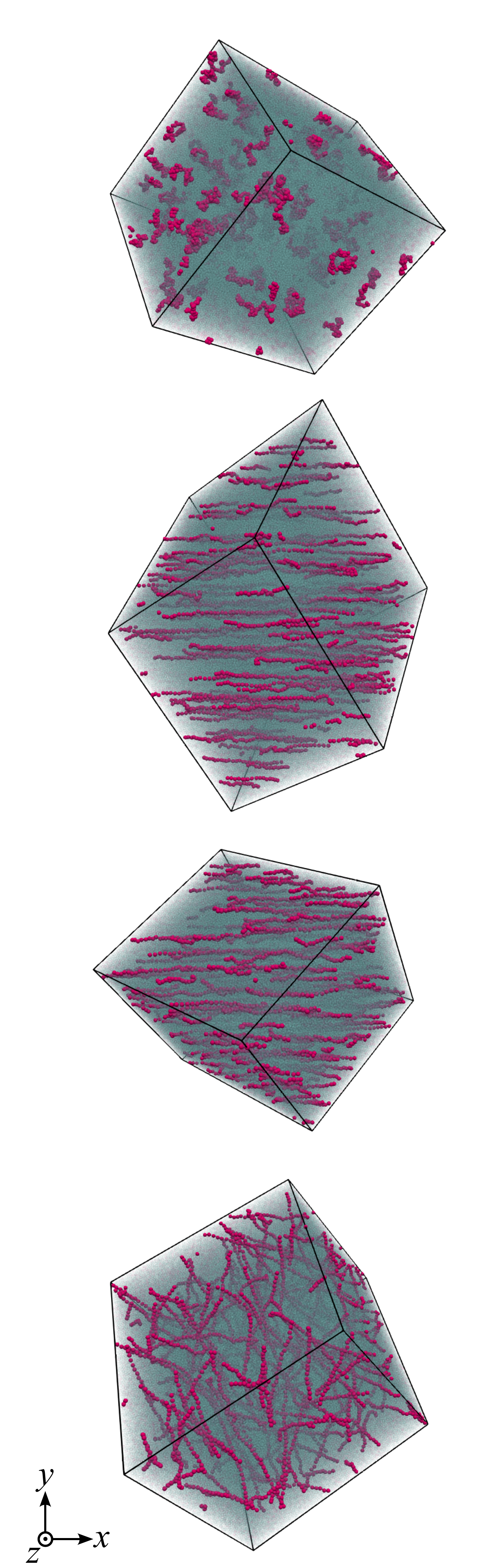} 
        \put(5,95){(a)}
        \put(5,70){(b)}
        \put(5,45){(c)}
        \put(5,23){(d)}
  \end{overpic}
    \caption{Snapshots of the polymer solution (a) at $t=0$ (i.e., just before the onset of the imposed flow) and (b-d) at $\dot{\epsilon}_\alpha t = 15$ in (b) uniaxial, (c) planar, and (d) biaxial extensional flows, with $\dot{\epsilon}_\alpha=0.0376$ for all cases. 
    Polymer and solvent particles are shown in red and blue, respectively. For clarity, all particles are rendered semi-transparent, except for selected polymers.}
  \label{fig:snapshot}
\end{figure}

We investigate three types of extensional flow.
Following previous studies~\cite{Koide2025-zr,koide2025relation}, we employ the SLLOD equations~\cite{evans_morriss_2008} and the generalized Kraynik--Reinelt~(GKR) boundary conditions~\cite{Dobson2014-kr,Hunt2016-vl} to generate homogeneous uniaxial, planar, and biaxial extensional flows.
The SLLOD equations are given by
\begin{align}
  &\frac{d{\bm{r}_i}}{dt} = {\bm{p}_i} + (\nabla \bm{u})^\mathsf{T} \cdot\bm{r}_i \\
  &\frac{d{\bm{p}}_i}{dt} = \bm{F}_i - (\nabla \bm{u})^\mathsf{T} \cdot\bm{p}_i ,
\label{eq:SLLOD_equation}
\end{align}
where $\bm{p}_i$ is the peculiar momentum of the $i$-th particle, $\bm{F}_i$ is the force acting on that particle, and $\nabla\bm{u}$ is the velocity gradient tensor of the imposed flow field.
For uniaxial, planar, and biaxial extensional flows, $\nabla\bm{u}$ is expressed as
\begin{align}
  \left(\nabla\bm{u}\right)_\mathrm{E} &=
  \begin{pmatrix}
      \dot{\epsilon} & 0 & 0\\
      0 & -\dot{\epsilon}/2 & 0\\
      0 & 0 & -\dot{\epsilon}/2
  \end{pmatrix},
  \label{eq:uniaxial}\\
  \left(\nabla\bm{u}\right)_\mathrm{P} &=
  \begin{pmatrix}
      \dot{\epsilon}_\mathrm{P} & 0 & 0\\
      0 & -\dot{\epsilon}_\mathrm{P} & 0\\
      0 & 0 & 0
  \end{pmatrix},
  \label{eq:planar}\\
  \left(\nabla\bm{u}\right)_\mathrm{B} &=
  \begin{pmatrix}
      \dot{\epsilon}_\mathrm{B} & 0 & 0\\
      0 & \dot{\epsilon}_\mathrm{B} & 0\\
      0 & 0 & -2\dot{\epsilon}_\mathrm{B}
  \end{pmatrix}.
  \label{eq:biaxial}
\end{align}
Here, $\dot{\epsilon}$ is the extension rate, $\dot{\epsilon}_\mathrm{P}$ is the planar extension rate, and $\dot{\epsilon}_\mathrm{B}$ is the biaxial extension rate. 
Hereafter, we denote these extension rates collectively by $\dot{\epsilon}_\alpha$.
Figure~\ref{fig:snapshot} shows snapshots of the polymer solution at $t=0$~(i.e., just before the onset of the imposed flow) and at $\dot{\epsilon}_\alpha t=15$ in the three extensional flows.
As shown in Fig.~\ref{fig:snapshot}, the simulation box deforms under the imposed flow.
Specifically, the deformation is governed by the evolution of the lattice basis vectors $\bm{e}_i$:
\begin{equation}
  \frac{d}{dt}\bm{e}_i = (\nabla \bm{u})^\mathsf{T} \cdot \bm{e}_i\label{eq:box_deform}.
\end{equation}
While applying extensional deformation, we systematically remap the deformed simulation box using the GKR method~\cite{Dobson2014-kr,Hunt2016-vl} with Semaev's algorithm~\cite{Semaev2001-sn}.
To prevent instabilities due to finite-precision arithmetic~\cite{Todd2000-rh}, the total peculiar momentum is periodically reset to zero, as in previous studies~\cite{Hunt2016-vl,Nicholson2016-aj}.

\begin{figure}
  \centering
  \begin{overpic}[width=0.7\linewidth]{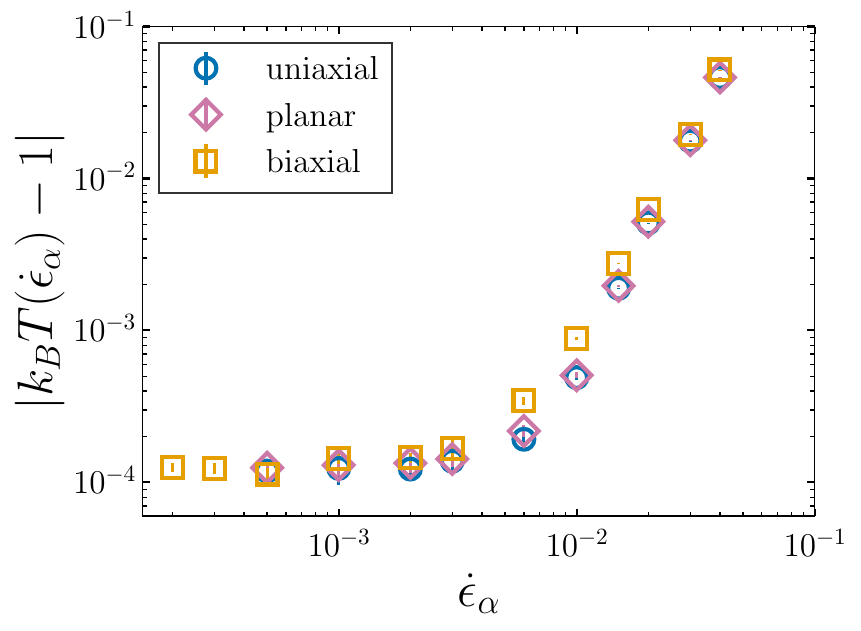} 
  \end{overpic}
    \caption{Relative temperature error $|k_BT(\dot{\epsilon}_\alpha)-1|$ as a function of the extension rate $\dot{\epsilon}_\alpha$ for uniaxial, planar, and biaxial extensional flows. The error bars denote the standard deviations from ten independent simulations.}
  \label{fig:temp}
\end{figure}

We describe the details of the numerical implementation.
The modified velocity Verlet method~\cite{Groot1997-je} is used for time integration, where the parameter $\lambda$ introduced in this scheme and the time step $\Delta t$ are set to $\lambda=0.65$ and $\Delta t=0.04$, respectively.
As shown in Fig.~\ref{fig:temp}, the relative temperature error $|k_BT(\dot{\epsilon}_\alpha)-1|$ increases with $\dot{\epsilon}_\alpha$, where $k_BT(\dot{\epsilon}_\alpha)$ is computed as $k_BT(\dot{\epsilon}_\alpha) = \langle \bm{p}^2\rangle/3$ with $\langle\cdot \rangle$ denoting the ensemble average.
Thus, we determine the range of $\dot{\epsilon}_\alpha$ so that $|k_BT(\dot{\epsilon}_\alpha)-1|<{0.06}$.
The initial simulation box is a cube with dimensions $60\times 60\times 60$.
First, equilibrium simulations are conducted from a random initial configuration for $12\,000$ time units, which is sufficiently long compared with the longest relaxation time $\tau(=160)$ of the polymers.
After this initial equilibration, we apply extensional flows to the polymer solutions.
All DPD simulations are conducted using our in-house code.

\section{Results and Discussion}\label{Results}
We first investigate the transient response of dilute polymer solutions under start-up extensional flows with different flow kinematics.
Uniaxial, planar, and biaxial extensional viscosity growth functions are defined, respectively, as 
\begin{align}
  &\eta_\mathrm{E}^+(t;\dot{\epsilon}) = \frac{\sigma_{xx}^+(t;\dot{\epsilon})-\{\sigma_{yy}^+(t;\dot{\epsilon})+\sigma_{zz}^+(t;\dot{\epsilon})\}/2}{\dot{\epsilon}},\label{eq:vis_uni} \\
  &\eta_\mathrm{P}^+(t;\dot{\epsilon}_\mathrm{P}) = \frac{\sigma_{xx}^+(t;\dot{\epsilon}_\mathrm{P})-\sigma_{yy}^+(t;\dot{\epsilon}_\mathrm{P})}{\dot{\epsilon}_\mathrm{P}}, \label{eq:vis_pla}\\
  &\eta_\mathrm{B}^+(t;\dot{\epsilon}_\mathrm{B}) = \frac{\{\sigma_{xx}^+(t;\dot{\epsilon}_\mathrm{B})+\sigma_{yy}^+(t;\dot{\epsilon}_\mathrm{B})\}/2-\sigma_{zz}^+(t;\dot{\epsilon}_\mathrm{B})}{\dot{\epsilon}_\mathrm{B}}. \label{eq:vis_bi}
\end{align}
Here, $\bm{\sigma}^+(t;\dot{\epsilon}_\alpha)$ is the stress tensor during start-up extensional flow, defined as~\cite{Irving1950-lc,Liu2015-yj}:
\begin{equation}
  \bm{\sigma}^+(t;\dot{\epsilon}_\alpha) = -\frac{1}{V}\left(\sum_{i<j} \bm{r}_{ij}(t)\bm{F}_{ij}(t)+\sum_{i} \bm{p}_{i}(t)\bm{p}_{i}(t)\right), 
\end{equation}
where $V$ is the system volume, $\bm{r}_{ij}(t)$ is the relative position vector of the $i$-th particle with respect to the $j$-th particle, $\bm{F}_{ij}(t)$ is the force exerted on the $i$-th particle by the $j$-th particle, and $\bm{p}_i(t)$ is the peculiar momentum of the $i$-th particle.
For simplicity, we denote these viscosity growth functions collectively by $\eta_{{\alpha}}^+(t;\dot{\epsilon}_\alpha)$.
To improve statistical accuracy, we evaluate $\eta_{{\alpha}}^+(t;\dot{\epsilon}_\alpha)$ by averaging the results from 20 independent simulations.
In addition, we apply a moving average with a window of $100\Delta t$ to reduce the noise.

\begin{figure*}
  \centering
      \begin{tabular}{c}
      \begin{minipage}{1\hsize}
        \centering
          \begin{overpic}[width=0.45\linewidth]{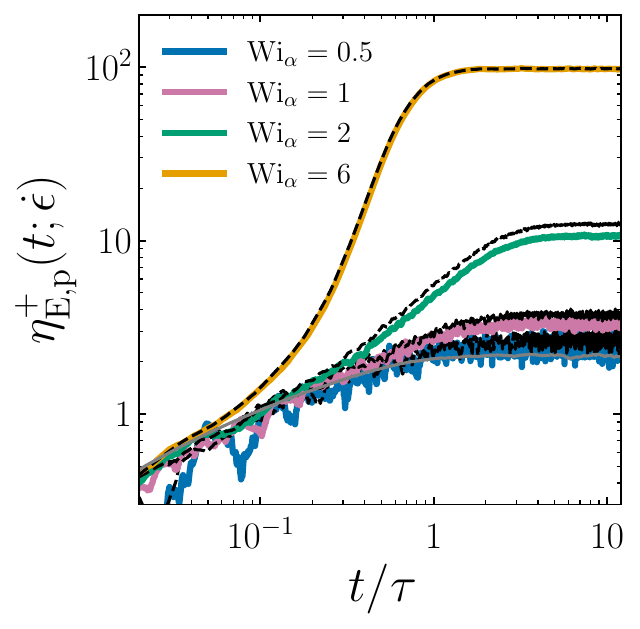}
              \linethickness{3pt}
        \put(1,90){(a)}

          \end{overpic}
      \end{minipage}\\
      \begin{minipage}{1\hsize}
        \centering

          \begin{overpic}[width=0.45\linewidth]{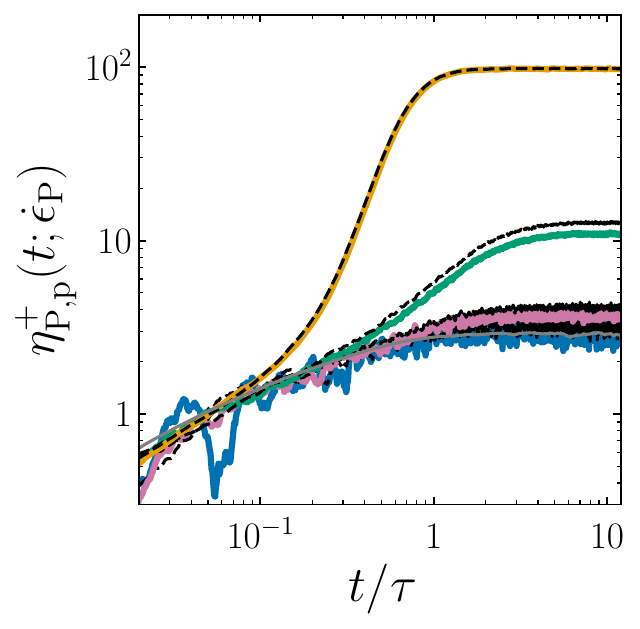}
              \linethickness{3pt}
        \put(1,90){(b)}

          \end{overpic}
      \end{minipage}\\
            \begin{minipage}{1\hsize}
        \centering
          \begin{overpic}[width=0.45\linewidth]{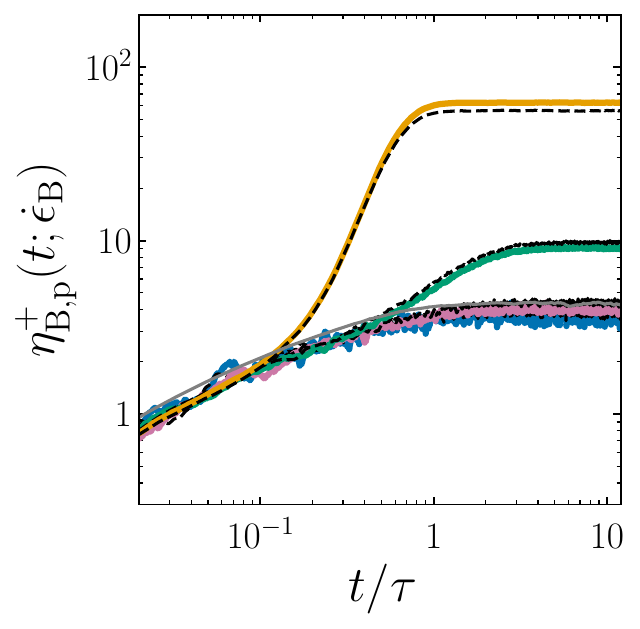}
              \linethickness{3pt}
        \put(1,90){(c)}

          \end{overpic}
      \end{minipage}
      \end{tabular}
    \caption{Polymer contribution $\eta_{\alpha,\mathrm{p}}^+(t;\dot{\epsilon}_\alpha)$~(thick solid curves) to the extensional viscosity growth function as a function of time $t$ normalized by the longest relaxation time $\tau$ of the polymers for (a) uniaxial, (b) planar, and (c) biaxial extensional flows at different Weissenberg numbers $\mathrm{Wi}_\alpha$. The thin gray curves represent the LVE envelopes. The black dashed curves represent the predictions based on the Rouse-type model~[Eq.~\eqref{eq:eta_Rouse_relation}]. }

      \label{fig:vis_growth}
\end{figure*}

We consider the polymer contribution $\eta_{\alpha,\mathrm{p}}^{+}(t;\dot{\epsilon}_\alpha)$ to $\eta_{{\alpha}}^+(t;\dot{\epsilon}_\alpha)$ by subtracting the Newtonian solvent contribution, namely $3\eta_\mathrm{s}$, $4\eta_\mathrm{s}$, and $6\eta_\mathrm{s}$ for uniaxial, planar, and biaxial extensional flows, respectively, where $\eta_\mathrm{s}$ is the solvent viscosity obtained from equilibrium DPD simulations of the solvent using the Green--Kubo formula~\cite{evans_morriss_2008}.
Figure~\ref{fig:vis_growth} shows $\eta_{\alpha,\mathrm{p}}^{+}(t;\dot{\epsilon}_\alpha)$ as a function of time $t$ normalized by the longest relaxation time $\tau$ of the polymers.
We evaluate $\tau$ by fitting the autocorrelation function $C(t)$ of the end-to-end vector to an exponential function $C(t)=C_0\exp(-t/\tau)$~\cite{Jiang2007-rf}.
The fitting range includes sufficiently long times, where higher-mode contributions are negligible for estimating $\tau$.
We show results at different values of the Weissenberg number $\mathrm{Wi}_\alpha=\tau\dot{\epsilon}_\alpha=0.5$, $1$, $2$, and $6$.
At $\mathrm{Wi}_\alpha=0.5$, $\eta_{\mathrm{E},\mathrm{p}}^{+}(t;\dot{\epsilon})$, $\eta_{\mathrm{P},\mathrm{p}}^{+}(t;\dot{\epsilon}_\mathrm{P})$, and $\eta_{\mathrm{B},\mathrm{p}}^{+}(t;\dot{\epsilon}_\mathrm{B})$ almost overlap with the linear viscoelastic~(LVE) envelopes $3\eta_{0,\mathrm{p}}^+(t)$, $4\eta_{0,\mathrm{p}}^+(t)$, and $6\eta_{0,\mathrm{p}}^+(t)$, respectively.
Here, we compute $\eta_{0,\mathrm{p}}^+(t)$ as
\begin{equation}
  \eta_{0,\mathrm{p}}^+(t) = \int_0^t G(t^\prime)dt^\prime - \eta_\mathrm{s}, 
\end{equation}
where $G(t)$ is the linear relaxation modulus obtained with the Green--Kubo formula.
For $\mathrm{Wi}_\alpha\gtrsim 1$, $\eta_{\alpha,\mathrm{p}}^{+}(t;\dot{\epsilon}_\alpha)$ initially follows the LVE envelope, whereas it exhibits an upward deviation at intermediate times, a phenomenon known as strain hardening, which has been observed experimentally in uniaxial extensional flows of polymer solutions~\cite{Sridhar1991-th,Gupta2000-ad,Anna2001-hj}.
As $\mathrm{Wi}_\alpha$ increases, the strain-hardening occurs at shorter times and becomes more evident.
Although strain hardening is observed in all extensional flows considered, it is relatively weak in biaxial extensional flows.
Specifically, at steady state for given $\mathrm{Wi}_\alpha(\gtrsim 1)$, $\eta_{\mathrm{B},\mathrm{p}}^{+}(t;\dot{\epsilon}_\mathrm{B})$ takes the smallest value, while $\eta_{\mathrm{E},\mathrm{p}}^{+}(t;\dot{\epsilon})\simeq \eta_{\mathrm{P},\mathrm{p}}^{+}(t;\dot{\epsilon}_\mathrm{P})$.
In the following, we aim to understand these behaviors of $\eta_{\alpha,\mathrm{p}}^{+}(t;\dot{\epsilon}_\alpha)$ by focusing on the flow-kinematics dependence of polymer conformation.

To relate the extensional viscosity growth function to polymer conformation, we adopt an analysis method based on a single-chain model, referred to as the Rouse-type model~\cite{uneyama2025radius}.
In the Rouse-type model, the original Rouse model~\cite{Rouse1953-hp} is extended to incorporate arbitrary interaction potentials between beads.
Specifically, each bead follows the overdamped Langevin equation
\begin{equation}
    \frac{d\bm{r}_i(t)}{dt} = (\nabla \bm{u})^\mathsf{T}\cdot \bm{r}_i(t)-\frac{1}{\zeta}\frac{\partial \mathcal{U}(\{\bm{r}_j(t)\})}{\partial \bm{r}_i(t)}+\sqrt{\frac{2k_BT}{\zeta}}\bm{w}_i(t),\label{eq:Rouse}
\end{equation}
where $\bm{r}_i(t)$ is the position of the $i$-th bead, $\zeta$ is the friction coefficient, $\mathcal{U}(\{\bm{r}_j(t)\})$ is the interaction potential for a single polymer chain, and $\bm{w}_i(t)$ is the Gaussian white-noise vector satisfying
\begin{align}
    &\langle \bm{w}_i(t)\rangle =0,\\
    &\langle \bm{w}_i(t)\bm{w}_j(s)\rangle =\delta_{ij}\delta(t-s)\bm{I},
\end{align}
where $\delta_{ij}$ is the Kronecker delta, $\delta(t)$ is the Dirac delta function, and $\bm{I}$ is the identity tensor.
Eq.~\eqref{eq:Rouse} can be viewed as a coarse-grained, approximate equation of motion for a single tagged polymer after eliminating the degrees of freedom of the surrounding solvent and other chains.
For the Rouse-type model, a relation between the stress tensor $\bm{\sigma}(t)$ and the polymer gyration tensor $\bm{S}(t)$ is derived~\cite{uneyama2025radius}: 
\begin{equation}
  \frac{d\bm{S}(t)}{dt} - (\nabla\bm{u})^\mathsf{T}\cdot \bm{S}(t)-\bm{S}(t)\cdot (\nabla\bm{u}) =-\frac{2}{\rho_\mathrm{p}\zeta}[\bm{\sigma}(t)-\bm{\sigma}_\mathrm{eq}], \label{eq:Rouse_relation}
\end{equation}
where $\rho_\mathrm{p}$ is the number density of beads, $\bm{\sigma}_\mathrm{eq}$ is the stress tensor at equilibrium and $\bm{S}(t)$ is defined as 
\begin{equation}
  \bm{S}(t) = \left\langle \frac{1}{N_\mathrm{p}}\sum_{i=1}^{N_\mathrm{p}}\{\bm{r}_{i}(t)-\bm{r}_{G}(t)\}\{\bm{r}_{i}(t)-\bm{r}_{G}(t)\}\right\rangle.
\end{equation}
Here, $N_\mathrm{p}$ is the number of beads in a chain and $\bm{r}_G(t)$ is the position of the center of mass of a chain.
Substituting $\nabla\bm{u}$ defined by Eqs.~\eqref{eq:uniaxial}-\eqref{eq:biaxial} into Eq.~\eqref{eq:Rouse_relation}, we obtain
\begin{equation}
  \eta_{\alpha,\mathrm{p}}^+(t;\dot{\epsilon}_\alpha) = \rho_\mathrm{p}\zeta \left[R_{g,+}^2(t;\dot{\epsilon}_\alpha)+\kappa_\alpha R_{g,-}^2(t;\dot{\epsilon}_\alpha)-\frac{1}{2\dot{\epsilon}_\alpha} \frac{d}{dt}\left\{R_{g,+}^2(t;\dot{\epsilon}_\alpha)-R_{g,-}^2(t;\dot{\epsilon}_\alpha)\right\}\right]\label{eq:eta_Rouse_relation},
\end{equation}
where $\kappa_\alpha$ represents the ratio of the magnitude of the compressional strain rate and the extensional strain rate~(i.e., $\kappa_\mathrm{E}=1/2$, $\kappa_\mathrm{P}=1$, and $\kappa_\mathrm{B}=2$).
For clarity, we denote the gyration radii in the extensional and compressional directions by $R_{g,+}(t;\dot{\epsilon}_\alpha)$ and $R_{g,-}(t;\dot{\epsilon}_\alpha)$, respectively.
For instance, in biaxial extensional flows, $R_{g,+}(t;\dot{\epsilon}_\mathrm{B})$ and $R_{g,-}(t;\dot{\epsilon}_\mathrm{B})$ are defined as 
\begin{align}
  &R_{g,+}^2(t;\dot{\epsilon}_\mathrm{B}) = \frac{1}{2}\left\langle \frac{1}{N_\mathrm{p}}\sum_{i=1}^{N_{\mathrm{p}}} \left[\{r_{i,x}(t)-r_{G,x}(t)\}^2+\{r_{i,y}(t)-r_{G,y}(t)\}^2\right]\right\rangle,\\
  &R_{g,-}^2(t;\dot{\epsilon}_\mathrm{B}) = \left\langle \frac{1}{N_\mathrm{p}}\sum_{i=1}^{N_{\mathrm{p}}} \{r_{i,z}(t)-r_{G,z}(t)\}^2\right\rangle.
\end{align}

We explore whether the relation~[Eq.~\eqref{eq:eta_Rouse_relation}] obtained from the Rouse-type model is applicable to our systems.
For this purpose, we estimate $\zeta$, which is an a priori unknown parameter, using the relation for the polymer contribution $\eta_{0,\mathrm{p}}$ to the zero-shear viscosity $\eta_{0}$ in the Rouse-type model~\cite{uneyama2025radius}:
\begin{equation}
  \eta_{0,\mathrm{p}} = \frac{\rho_\mathrm{p}\zeta }{6}R_{g,\mathrm{eq}}^2,\label{eq:zeroshear}
\end{equation}
where $R_{g,\mathrm{eq}}$ is the gyration radius at equilibrium.
We calculate $\zeta$ by substituting $R_{g,\mathrm{eq}}$ and $\eta_{0,\mathrm{p}}$ obtained from equilibrium DPD simulations into Eq.~\eqref{eq:zeroshear}.
In Fig.~\ref{fig:vis_growth}, we show the predictions based on the Rouse-type model computed from the right-hand side of Eq.~\eqref{eq:eta_Rouse_relation}~(black dashed curves).
They are smoothed using the same moving average as $\eta_{\alpha,\mathrm{p}}^{+}(t;\dot{\epsilon}_\alpha)$.
These predictions quantitatively reproduce $\eta_{\alpha,\mathrm{p}}^{+}(t;\dot{\epsilon}_\alpha)$ obtained with DPD simulations.
Although a slight discrepancy is observed at steady state depending on $\mathrm{Wi}_\alpha$, which we discuss later in detail, we demonstrate that the Rouse-type model enables us to analyze $\eta_{\alpha,\mathrm{p}}^{+}(t;\dot{\epsilon}_\alpha)$ in terms of the polymer gyration radii $R_{g,+}(t;\dot{\epsilon}_\alpha)$ and $R_{g,-}(t;\dot{\epsilon}_\alpha)$.

\begin{figure*}
  \centering
      \begin{tabular}{llll}
      \begin{minipage}{0.25\hsize}
        \centering
          \begin{overpic}[width=1\linewidth]{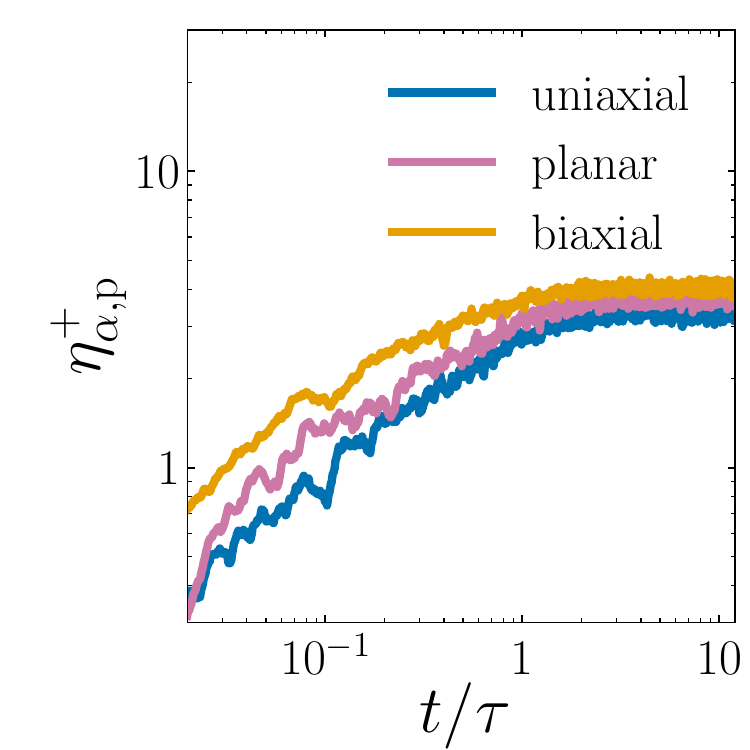}
              \linethickness{3pt}
        \put(-1,100){(a)}

          \end{overpic}
      \end{minipage}
          \begin{minipage}{0.25\hsize}
          \begin{overpic}[width=1\linewidth]{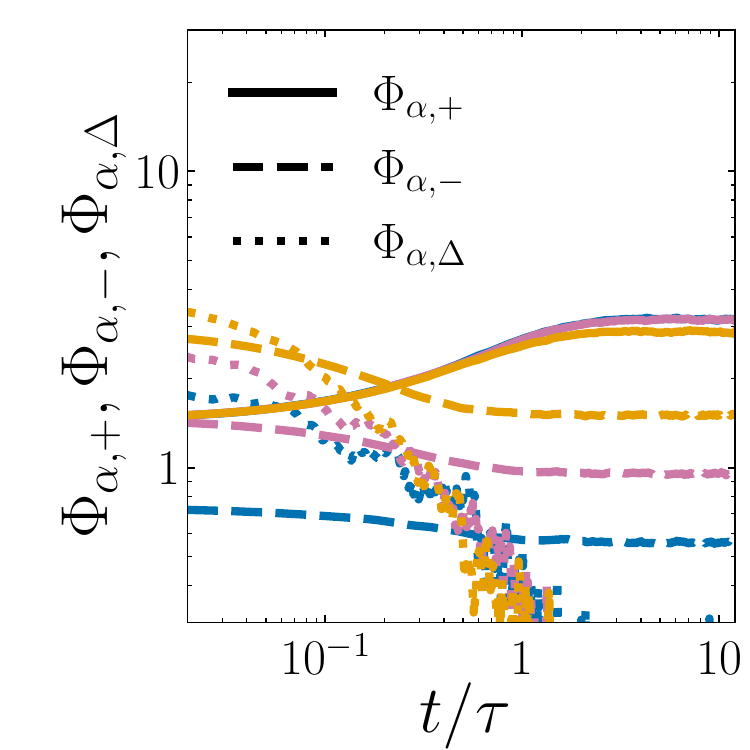}
              \linethickness{3pt}
        \put(-1,100){(b)}

          \end{overpic}
      \end{minipage}
      \begin{minipage}{0.25\hsize}
          \begin{overpic}[width=1\linewidth]{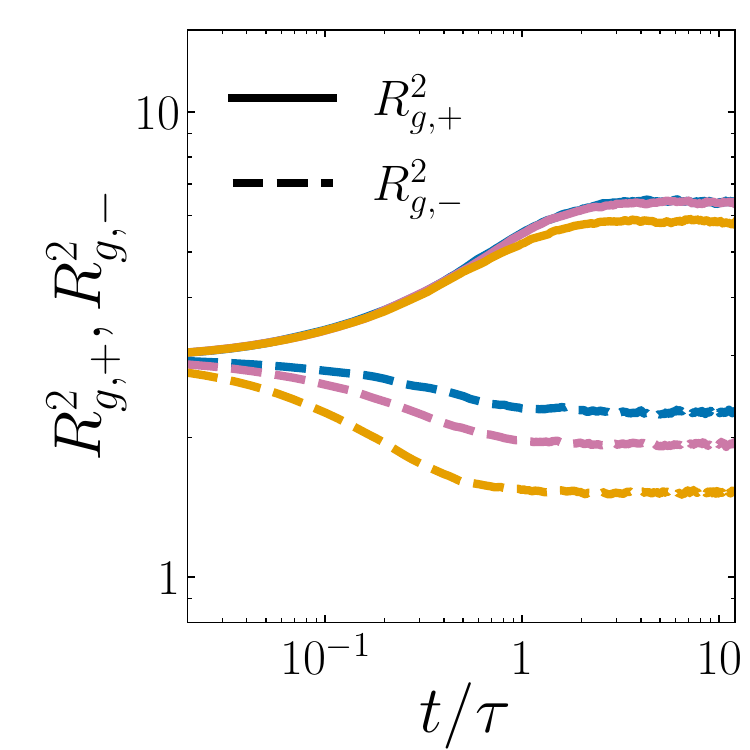}
              \linethickness{3pt}
        \put(-1,100){(c)}

          \end{overpic}
      \end{minipage}
            \begin{minipage}{0.25\hsize}
          \begin{overpic}[width=1\linewidth]{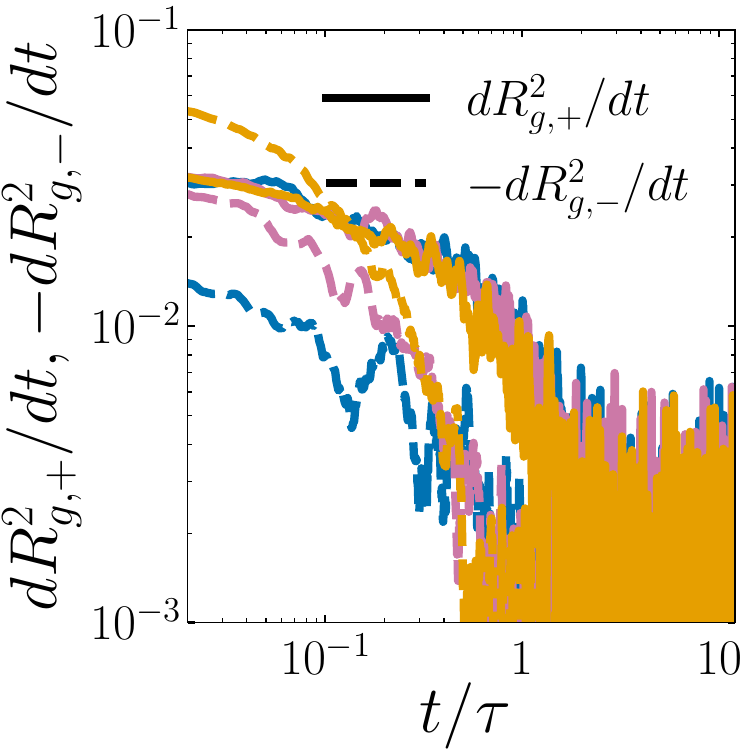}
              \linethickness{3pt}
        \put(-1,100){(d)}

          \end{overpic}
      \end{minipage}\\
      \begin{minipage}{0.25\hsize}
          \begin{overpic}[width=1\linewidth]{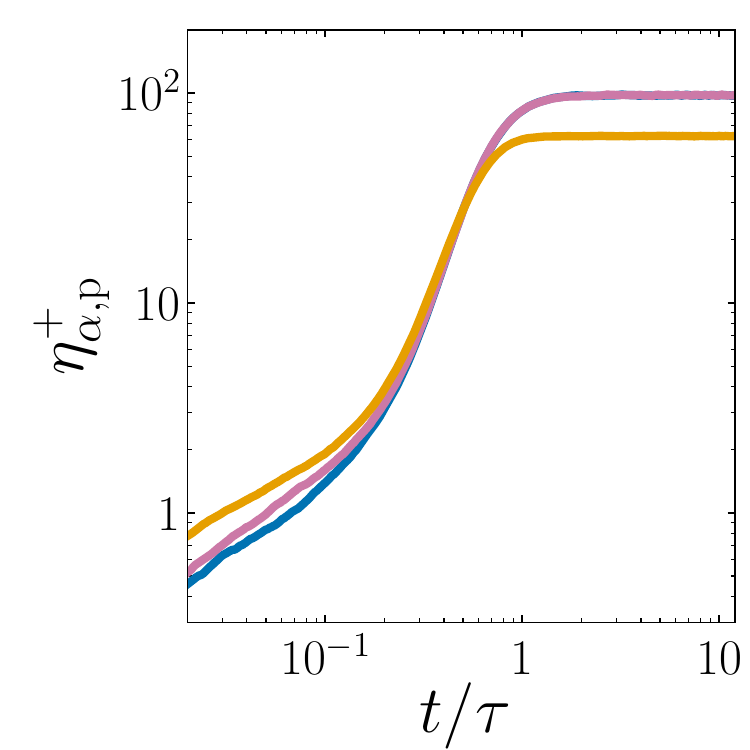}
              \linethickness{3pt}
        \put(-1,100){(e)}

          \end{overpic}
      \end{minipage}
          \begin{minipage}{0.25\hsize}
          \begin{overpic}[width=1\linewidth]{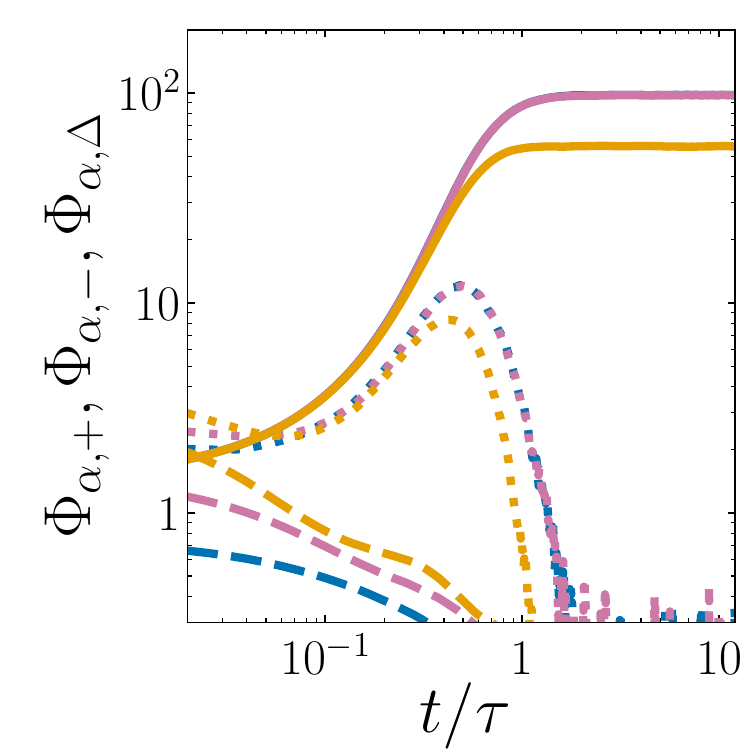}
              \linethickness{3pt}
        \put(-1,100){(f)}

          \end{overpic}
      \end{minipage}
      \begin{minipage}{0.25\hsize}
          \begin{overpic}[width=1\linewidth]{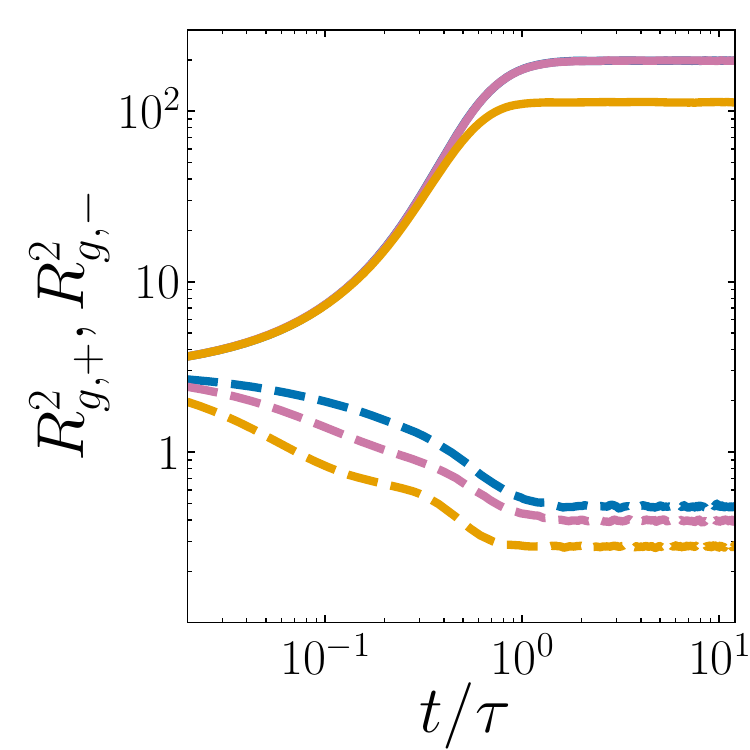}
              \linethickness{3pt}
        \put(-1,100){(g)}

          \end{overpic}
      \end{minipage}
            \begin{minipage}{0.25\hsize}
          \begin{overpic}[width=1\linewidth]{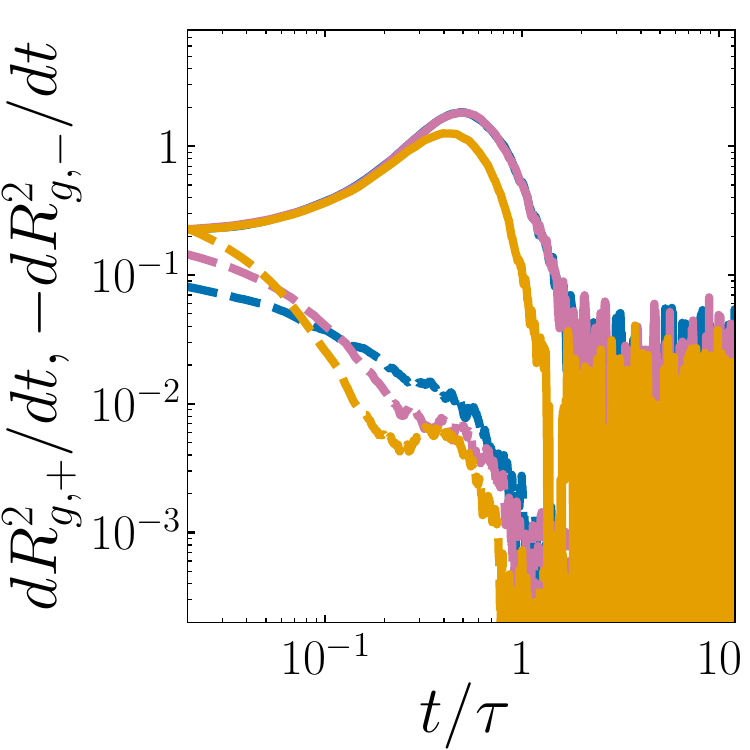}
              \linethickness{3pt}
        \put(-1,100){(h)}

          \end{overpic}
      \end{minipage}

      \end{tabular}
  \caption{Time series for (a-d)~$\mathrm{Wi}_\alpha=1$ and (e-h)~$\mathrm{Wi}_\alpha=6$: (a,e) $\eta_{\alpha,\mathrm{p}}^{+}(t;\dot{\epsilon}_\alpha)$; (b,f) $\Phi_{\alpha,+}(t;\dot{\epsilon}_\alpha)$, $\Phi_{\alpha,-}(t;\dot{\epsilon}_\alpha)$, and $\Phi_{\alpha,\Delta}(t;\dot{\epsilon}_\alpha)$; (c,g) $R_{g,+}^2(t;\dot{\epsilon}_\alpha)$ and $R_{g,-}^2(t;\dot{\epsilon}_\alpha)$; (d,h) ${d R_{g,+}^2(t;\dot{\epsilon}_\alpha)}/{dt}$ and $-dR_{g,-}^2(t;\dot{\epsilon}_\alpha)/dt$.}

      \label{fig:vis_growth_gyration}
\end{figure*}

\begin{figure}
  \centering
  \begin{overpic}[width=0.7\linewidth]{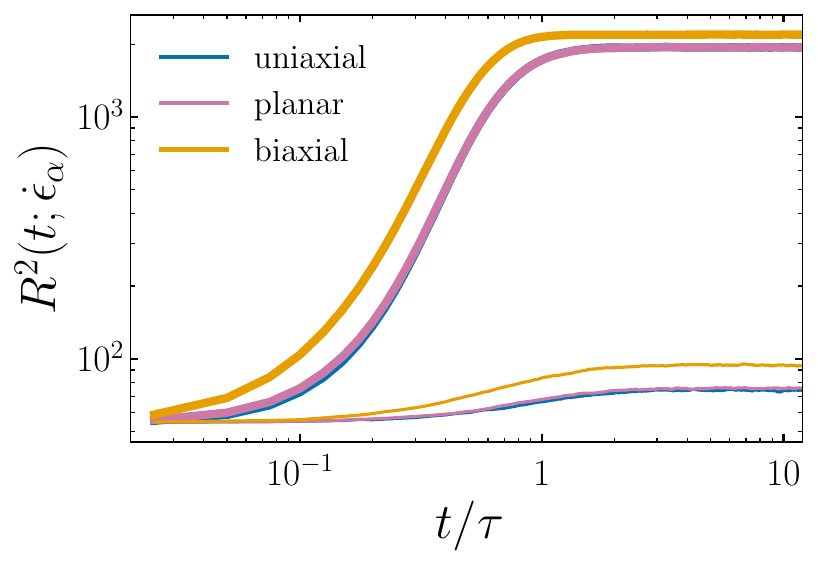} 
  \end{overpic}
  \caption{Mean-square end-to-end distance $R^2(t;\dot{\epsilon}_\alpha)$ as a function of $t/\tau$ at $\mathrm{Wi}_\alpha=1$~(thin curves) and $6$~(thick curves).}
  \label{fig:endtoend}
\end{figure}

Using the Rouse-type model, we explore how flow kinematics affects the extensional viscosity growth function in terms of the conformational changes of polymers.
We focus on the contribution from each term in the right-hand side of Eq.~\eqref{eq:eta_Rouse_relation}.
Specifically, we decompose $\eta_{\alpha,\mathrm{p}}^{+}(t;\dot{\epsilon}_\alpha)$ as $\eta_{\alpha,\mathrm{p}}^{+}(t;\dot{\epsilon}_\alpha)=\Phi_{\alpha,+}(t;\dot{\epsilon}_\alpha)+\Phi_{\alpha,-}(t;\dot{\epsilon}_\alpha)-\Phi_{\alpha,\Delta}(t;\dot{\epsilon}_\alpha)$, where $\Phi_{\alpha,+}(t;\dot{\epsilon}_\alpha)$, $\Phi_{\alpha,-}(t;\dot{\epsilon}_\alpha)$, and $\Phi_{\alpha,\Delta}(t;\dot{\epsilon}_\alpha)$ are defined as
\begin{align}
  & \Phi_{\alpha,+}(t;\dot{\epsilon}_\alpha) = \rho_\mathrm{p}\zeta R_{g,+}^2(t;\dot{\epsilon}_\alpha), \label{eq:Phi_plus}\\
  & \Phi_{\alpha,-}(t;\dot{\epsilon}_\alpha) = \kappa_\alpha\rho_\mathrm{p}\zeta R_{g,-}^2(t;\dot{\epsilon}_\alpha), \label{eq:Phi_minus}\\
  & \Phi_{\alpha,\Delta}(t;\dot{\epsilon}_\alpha) = \frac{\rho_\mathrm{p}\zeta}{2\dot{\epsilon}_\alpha} \frac{d}{dt}\left\{R_{g,+}^2(t;\dot{\epsilon}_\alpha)-R_{g,-}^2(t;\dot{\epsilon}_\alpha)\right\}. \label{eq:Phi_delta}
\end{align}
To investigate the contributions in different extensional flows at $\mathrm{Wi}_\alpha=1$, Fig.~\ref{fig:vis_growth_gyration}(a) shows $\eta_{\alpha,\mathrm{p}}^{+}(t;\dot{\epsilon}_\alpha)$, and Fig.~\ref{fig:vis_growth_gyration}(b) shows $\Phi_{\alpha,+}(t;\dot{\epsilon}_\alpha)$, $\Phi_{\alpha,-}(t;\dot{\epsilon}_\alpha)$, and $\Phi_{\alpha,\Delta}(t;\dot{\epsilon}_\alpha)$.
At $\mathrm{Wi}_\alpha=1$, the relation $\eta_{\mathrm{B},\mathrm{p}}^{+}(t;\dot{\epsilon}_\mathrm{B})>\eta_{\mathrm{P},\mathrm{p}}^{+}(t;\dot{\epsilon}_\mathrm{P})>\eta_{\mathrm{E},\mathrm{p}}^{+}(t;\dot{\epsilon})$ holds over the entire time range.
In the growth regime of $\eta_{\alpha,\mathrm{p}}^{+}(t;\dot{\epsilon}_\alpha)$, $\Phi_{\alpha,+}(t;\dot{\epsilon}_\alpha)$, $\Phi_{\alpha,-}(t;\dot{\epsilon}_\alpha)$, and $\Phi_{\alpha,\Delta}(t;\dot{\epsilon}_\alpha)$ exhibit comparable contributions to $\eta_{\alpha,\mathrm{p}}^{+}(t;\dot{\epsilon}_\alpha)$.
As $t/\tau$ increases, $\Phi_{\alpha,+}(t;\dot{\epsilon}_\alpha)$ increases, $\Phi_{\alpha,-}(t;\dot{\epsilon}_\alpha)$ decreases, and $\Phi_{\alpha,\Delta}(t;\dot{\epsilon}_\alpha)$ approaches $0$.
To relate these rheological responses to polymer conformation, we also show $R_{g,+}^2(t;\dot{\epsilon}_\alpha)$ and $R_{g,-}^2(t;\dot{\epsilon}_\alpha)$ in Fig.~\ref{fig:vis_growth_gyration}(c) and ${dR_{g,+}^2(t;\dot{\epsilon}_\alpha)}/{dt}$ and $-dR_{g,-}^2(t;\dot{\epsilon}_\alpha)/dt$ in Fig.~\ref{fig:vis_growth_gyration}(d).
For $t/\tau\lesssim 1$, $R_{g,+}^2(t;\dot{\epsilon}_\alpha)$ is almost independent of the flow type because the strain rate in the extensional direction is common in these extensional flows, and the coupling between components is very weak.
In contrast, $R_{g,-}^2(t;\dot{\epsilon}_\alpha)$ decreases most in biaxial extensional flow and least in uniaxial extensional flow, consistent with the magnitude of the compressional strain rate.
Accordingly, $-dR_{g,-}^2(t;\dot{\epsilon}_\alpha)/dt$ takes the largest value in biaxial extensional flow, while ${d R_{g,+}^2(t;\dot{\epsilon}_\alpha)}/{dt}$ is independent of the flow type.
Although polymer compression reduces $\eta_{\alpha,\mathrm{p}}^{+}(t;\dot{\epsilon}_\alpha)$~(see Eq.~\eqref{eq:eta_Rouse_relation}) and is strongest in biaxial extensional flow, $\kappa_\alpha$ in Eq.~\eqref{eq:Phi_minus} is also largest in this case.
Thus, the relation $\eta_{\mathrm{B},\mathrm{p}}^{+}(t;\dot{\epsilon}_\mathrm{B})>\eta_{\mathrm{P},\mathrm{p}}^{+}(t;\dot{\epsilon}_\mathrm{P})>\eta_{\mathrm{E},\mathrm{p}}^{+}(t;\dot{\epsilon})$ at $\mathrm{Wi}_\alpha=1$ mainly arises from the component $\Phi_{\alpha,-}(t;\dot{\epsilon}_\alpha)$ associated with the compressional direction, especially through $\kappa_\alpha$, which is determined solely by the flow kinematics.
Figure~\ref{fig:vis_growth_gyration}(e--h) shows the results at $\mathrm{Wi}_\alpha=6$.
Although the initial response of $\eta_{\alpha,\mathrm{p}}^{+}(t;\dot{\epsilon}_\alpha)$ is similar to that at $\mathrm{Wi}_\alpha=1$, $\eta_{\alpha,\mathrm{p}}^{+}(t;\dot{\epsilon}_\alpha)$ in different extensional flows nearly collapse for $0.3\lesssim t/\tau\lesssim 0.5$, and the relation $\eta_{\mathrm{E},\mathrm{p}}^{+}(t;\dot{\epsilon})\simeq \eta_{\mathrm{P},\mathrm{p}}^{+}(t;\dot{\epsilon}_\mathrm{P})>\eta_{\mathrm{B},\mathrm{p}}^{+}(t;\dot{\epsilon}_\mathrm{B})$ holds at steady state~[Fig.~\ref{fig:vis_growth_gyration}(e)].
Figure~\ref{fig:vis_growth_gyration}(g) demonstrates that at $\mathrm{Wi}_\alpha=6$, $R_{g,+}^2(t;\dot{\epsilon}_\alpha)$ becomes nearly two orders of magnitude larger than $R_{g,-}^2(t;\dot{\epsilon}_\alpha)$, thus leading to $\Phi_{\alpha,+}(t;\dot{\epsilon}_\alpha)$ being dominant in Fig.~\ref{fig:vis_growth_gyration}(f).
In addition, $R_{g,+}^2(t;\dot{\epsilon}_\mathrm{B})$ in biaxial extensional flow becomes smaller than that in the other flows for $t/\tau\gtrsim 0.5$.
Since biaxial extensional flow has two extensional directions, the stretching degree in each direction becomes smaller as polymers approach a fully extended state.
Consequently, at $\mathrm{Wi}_\alpha=6$, where nonlinear effects are nonnegligible, $\eta_{\mathrm{B},\mathrm{p}}^{+}(t;\dot{\epsilon}_\mathrm{B})$ eventually takes the smallest value.
Incidentally, although $R_{g,+}^2(t;\dot{\epsilon}_\alpha)$, which is essentially related to $\eta_{\alpha,\mathrm{p}}^{+}(t;\dot{\epsilon}_\alpha)$, becomes the smallest in biaxial extensional flow, Fig.~\ref{fig:endtoend} demonstrates that the mean-square end-to-end distance $R^2(t;\dot{\epsilon}_\alpha)=\langle|\bm{r}_{N_\mathrm{p}}(t)-\bm{r}_1(t)|^2\rangle$ becomes the largest under biaxial extensional flow at $\mathrm{Wi}_\alpha$ considered.


\begin{figure}
  \centering
  \begin{overpic}[width=0.7\linewidth]{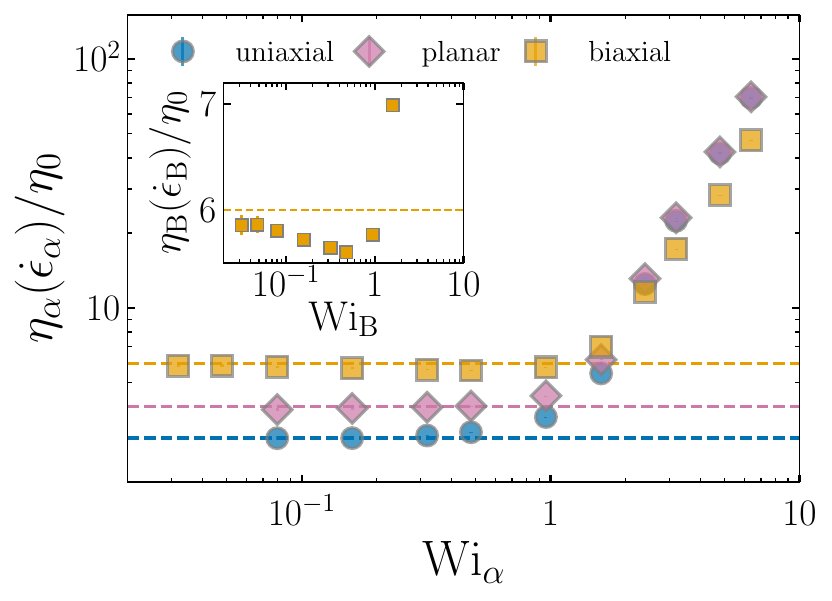} 
  \end{overpic}
  \caption{Steady-state extensional viscosity $\eta_\alpha(\dot{\epsilon}_\alpha)$ normalized by the zero-shear viscosity $\eta_0$ as a function of the Weissenberg number $\mathrm{Wi}_\alpha$. 
  The dashed lines indicate the corresponding Trouton ratios predicted by linear viscoelasticity: $\eta_\mathrm{E}(\dot{\epsilon})/\eta_0=3$, $\eta_\mathrm{P}(\dot{\epsilon}_\mathrm{P})/\eta_0=4$, and $\eta_\mathrm{B}(\dot{\epsilon}_\mathrm{B})/\eta_0=6$. The inset shows the close-up view of $\eta_\mathrm{B}(\dot{\epsilon}_\mathrm{B})/\eta_0$. The error bars denote the standard deviations from ten independent simulations.}

  \label{fig:steady_vis}
\end{figure}

We then investigate the effect of flow kinematics on the steady-state extensional viscosity $\eta_\alpha(\dot{\epsilon}_\alpha)$ of dilute polymer solutions.
We evaluate $\eta_\alpha(\dot{\epsilon}_\alpha)$ by time-averaging $\eta_{\alpha}^{+}(t;\dot{\epsilon}_\alpha)$ over $t/\tau\gtrsim 10$.
Figure~\ref{fig:steady_vis} shows $\eta_\alpha(\dot{\epsilon}_\alpha)$ normalized by $\eta_0$ as a function of $\mathrm{Wi}_\alpha$.
For $\mathrm{Wi}_\alpha\ll 1$, $\eta_\alpha(\dot{\epsilon}_\alpha)/\eta_0$ approaches $3$, $4$, and $6$ in uniaxial, planar, and biaxial extensional flows, respectively, consistent with the Trouton ratios predicted by linear viscoelasticity~\cite{Bird1987DPLv1}.
For $\mathrm{Wi}_\alpha\gtrsim 1$, $\eta_\alpha(\dot{\epsilon}_\alpha)$ increases with $\mathrm{Wi}_\alpha$, while $\eta_\mathrm{B}(\dot{\epsilon}_\mathrm{B})$ shows a slight decrease for $0.1\lesssim \mathrm{Wi}_\mathrm{B}\lesssim 1$~(see the inset of Fig.~\ref{fig:steady_vis}).
In addition, as $\mathrm{Wi}_\alpha$ increases, the relation $\eta_\mathrm{E}(\dot{\epsilon})\simeq \eta_\mathrm{P}(\dot{\epsilon}_\mathrm{P})>\eta_\mathrm{B}(\dot{\epsilon}_\mathrm{B})$ becomes apparent, which is consistent with the recent experimental results~\cite{Haward2023b}.
It is also worth noting the Weissenberg number at which $\eta_\mathrm{\alpha}(\dot{\epsilon}_\alpha)$ begins to increase noticeably.
For the FENE dumbbell model, $\eta_\mathrm{\alpha}(\dot{\epsilon}_\alpha)$ exhibits a sharp increase around $\tau_H\dot{\epsilon}_\alpha\simeq 0.5$~\cite{bird1987dynamics}, where $\tau_H$ denotes the stress relaxation time in the Hookean limit.
Although this may appear inconsistent with our results, where $\eta_\mathrm{\alpha}(\dot{\epsilon}_\alpha)$ begins to increase around $\mathrm{Wi}_\alpha\simeq 1$, the discrepancy can be partly attributed to the definition of the Weissenberg number.
Indeed, in the Rouse model, the stress relaxation time is half the conformational relaxation time~\cite{rubinstein2003polymer}.
The limiting behavior of $\eta_\mathrm{\alpha}(\dot{\epsilon}_\alpha)$ as $\mathrm{Wi}_\alpha \to \infty$ is also important for comparison with other models.
However, the plateau region of $\eta_\alpha(\dot{\epsilon}_\alpha)$ is not reached in Fig.~\ref{fig:steady_vis}, because our results are limited to $\mathrm{Wi}_\alpha\lesssim 10$ due to the temperature rise at high extension rates~(Fig.~\ref{fig:temp}).

\begin{figure*}
  \centering
      \begin{tabular}{c}
      \begin{minipage}{1\hsize}
        \centering
          \begin{overpic}[width=0.45\linewidth]{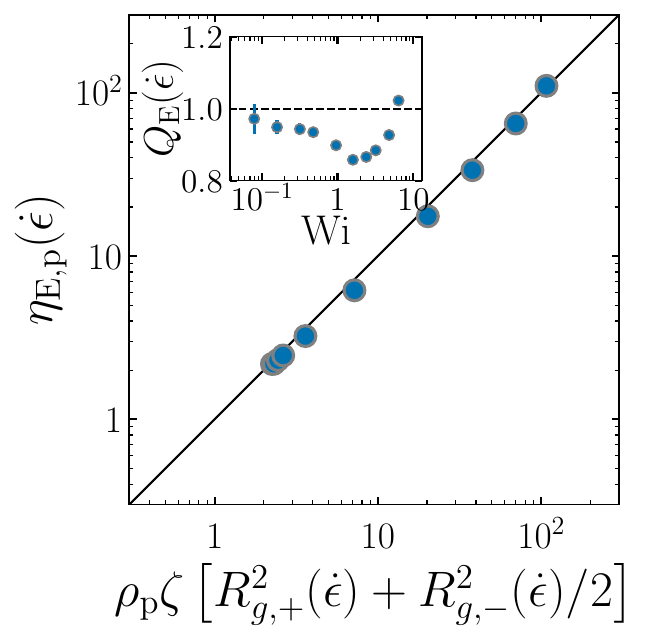}
              \linethickness{3pt}
        \put(1,90){(a)}

          \end{overpic}
      \end{minipage}\\
      \begin{minipage}{1\hsize}
        \centering
          \begin{overpic}[width=0.45\linewidth]{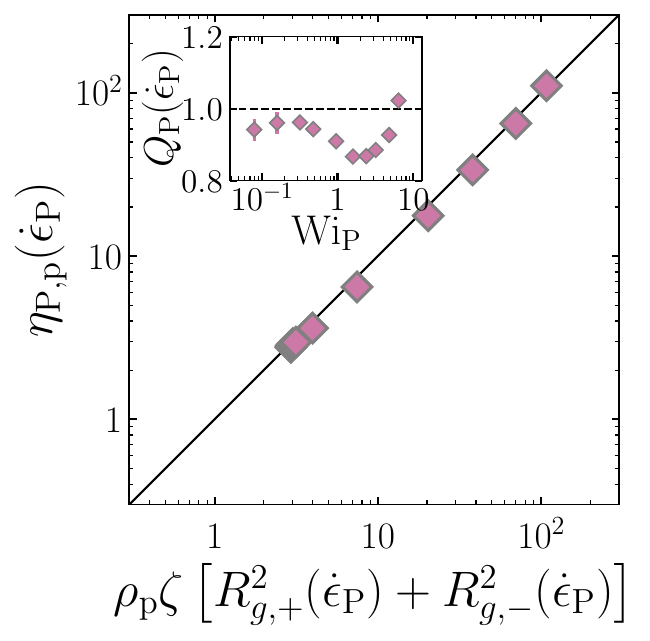}
              \linethickness{3pt}
        \put(1,90){(b)}

          \end{overpic}
      \end{minipage}\\
            \begin{minipage}{1\hsize}
        \centering
          \begin{overpic}[width=0.45\linewidth]{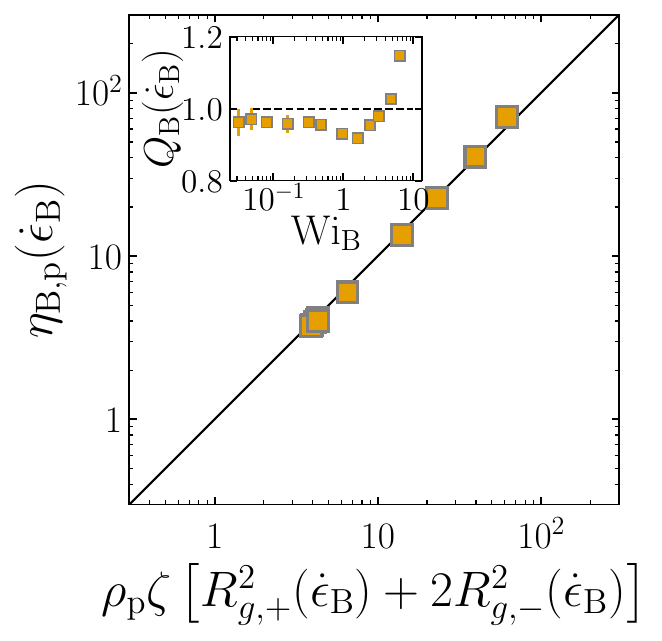}
              \linethickness{3pt}
        \put(1,90){(c)}

          \end{overpic}
      \end{minipage}
      \end{tabular}
        \caption{Polymer contribution $\eta_{\alpha,\mathrm{p}}(\dot{\epsilon}_\alpha)$ to the steady-state extensional viscosity $\eta_\alpha(\dot{\epsilon}_\alpha)$ as a function of $\rho_\mathrm{p} \zeta [R_{g,+}^2(\dot{\epsilon}_\alpha)+\kappa_\alpha R_{g,-}^2(\dot{\epsilon}_\alpha)]$ for (a) uniaxial, (b) planar, and (c) biaxial extensional flows. The black solid lines indicate $\eta_{\alpha,\mathrm{p}}(\dot{\epsilon}_\alpha) = \rho_\mathrm{p}\zeta [R_{g,+}^2(\dot{\epsilon}_\alpha)+\kappa_\alpha R_{g,-}^2(\dot{\epsilon}_\alpha)]$. The insets in each panel show $Q_\alpha(\dot{\epsilon}_\alpha)$ as a function of $\mathrm{Wi}_\alpha$, where the black dashed lines indicate $Q_\alpha(\dot{\epsilon}_\alpha)=1$. The error bars denote the standard deviations from ten independent simulations.}

      \label{fig:steady_vis_gyration}
\end{figure*}

We aim to explain the flow-kinematics dependence of $\eta_\alpha(\dot{\epsilon}_\alpha)$ in terms of polymer conformation.
The Rouse-type model provides a quantitative relation between the steady-state extensional viscosity and the polymer gyration radius.
At steady state, Eq.~\eqref{eq:eta_Rouse_relation} reduces to 
\begin{equation}
  \eta_{\alpha,\mathrm{p}}(\dot{\epsilon}_\alpha) = \rho_\mathrm{p}\zeta \left[ R_{g,+}^2(\dot{\epsilon}_\alpha)+\kappa_\alpha R_{g,-}^2(\dot{\epsilon}_\alpha)\right],\label{eq:eta_Rouse_relation_steady}
\end{equation}
where $\eta_{\alpha,\mathrm{p}}(\dot{\epsilon}_\alpha)$ is the polymer contribution to $\eta_{\alpha}(\dot{\epsilon}_\alpha)$.
To verify the validity of Eq.~\eqref{eq:eta_Rouse_relation_steady}, we show $ \eta_{\alpha,\mathrm{p}}(\dot{\epsilon}_\alpha)$ as a function of $\rho_\mathrm{p}\zeta \left[ R_{g,+}^2(\dot{\epsilon}_\alpha)+\kappa_\alpha R_{g,-}^2(\dot{\epsilon}_\alpha)\right]$ in Fig.~\ref{fig:steady_vis_gyration}.
For each extensional flow, $\eta_{\alpha,\mathrm{p}}(\dot{\epsilon}_\alpha)$ is in good agreement with $\rho_\mathrm{p}\zeta \left[ R_{g,+}^2(\dot{\epsilon}_\alpha)+\kappa_\alpha R_{g,-}^2(\dot{\epsilon}_\alpha)\right]$, including results under uniaxial extensional flow reported in our previous work~\cite{koide2025relation}.
For a more quantitative analysis, the inset of Fig.~\ref{fig:steady_vis_gyration} shows the ratio $Q_\alpha(\dot{\epsilon}_\alpha)$ of the left-hand side to the right-hand side of Eq.~\eqref{eq:eta_Rouse_relation_steady}, defined as $Q_\alpha(\dot{\epsilon}_\alpha) =\eta_{\alpha,\mathrm{p}}(\dot{\epsilon}_\alpha)/\{\rho_\mathrm{p}\zeta \left[ R_{g,+}^2(\dot{\epsilon}_\alpha)+\kappa_\alpha R_{g,-}^2(\dot{\epsilon}_\alpha)\right]\}$, as a function of $\mathrm{Wi}_\alpha$.
Overall, Eq.~\eqref{eq:eta_Rouse_relation_steady} holds within $\pm 20\%$ irrespective of flow type and $\mathrm{Wi}_\alpha$.
However, $Q_\alpha(\dot{\epsilon}_\alpha)$ exhibits a nonmonotonic deviation from unity as $\mathrm{Wi}_\alpha$ increases, with a weak dependence on flow type.
While we assume that the friction coefficient $\zeta$ is constant for given $N_\mathrm{p}$ and $\phi$, the observed deviation of $Q_\alpha(\dot{\epsilon}_\alpha)$ from unity may be interpreted as the modulation of $\zeta$ due to extensional flows.
Indeed, flow-induced changes in the effective friction coefficient have been extensively discussed as one of the key factors underlying nonlinear responses under fast flows in both dilute~\cite{Larson2005-ef,Prabhakar2016-ba} and concentrated systems~\cite{Ianniruberto2020-rg,Matsumiya2021-ti}.
Refining our modeling approach to incorporate $\dot{\epsilon}_\alpha$-dependent $\zeta$ is left for future work.

\begin{figure}
  \centering
  \begin{overpic}[width=0.5\linewidth]{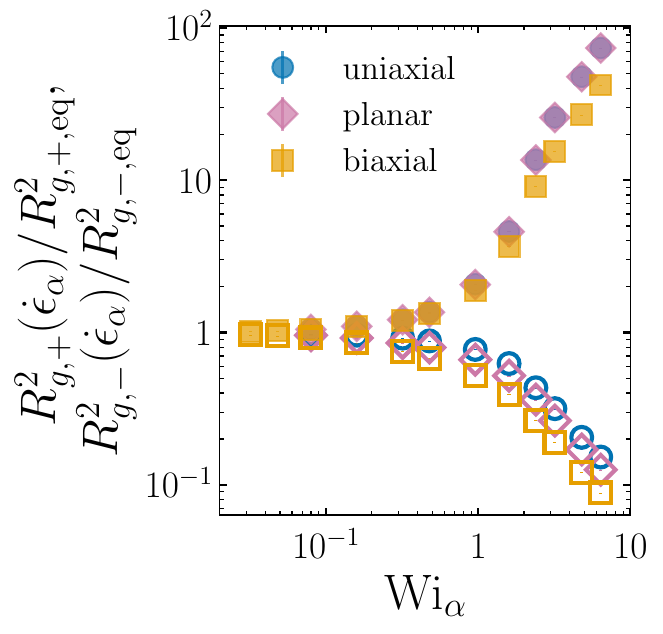} 
  \end{overpic}
  \caption{Normalized mean-square gyration radii of polymers in the extensional $R_{g,+}^2(\dot{\epsilon}_\alpha)/R_{g,+,\mathrm{eq}}^2$~(filled symbols) and compressional $R_{g,-}^2(\dot{\epsilon}_\alpha)/R_{g,-,\mathrm{eq}}^2$~(open symbols) directions as functions of the Weissenberg number $\mathrm{Wi}_\alpha$ for uniaxial, planar, and biaxial extensional flows.
  The error bars denote the standard deviations from ten independent simulations and are smaller than the symbol size.}

  \label{fig:gyration}
\end{figure}

On the basis of the established relation~(Fig.~\ref{fig:steady_vis_gyration}), we elucidate the physical origin of the flow-kinematics dependence of the steady-state extensional viscosity from the viewpoint of polymer conformation.
We first examine how the gyration radius depends on the flow type and the Weissenberg number.
Figure~\ref{fig:gyration} shows the mean-square gyration radii $R_{g,+}^2(\dot{\epsilon}_\alpha)$ and $R_{g,-}^2(\dot{\epsilon}_\alpha)$ in the extensional and compressional directions as functions of $\mathrm{Wi}_\alpha$.
Here, both $R_{g,+}^2(\dot{\epsilon}_\alpha)$ and $R_{g,-}^2(\dot{\epsilon}_\alpha)$ are normalized by the corresponding equilibrium values $R_{g,+,\mathrm{eq}}^2$ and $R_{g,-,\mathrm{eq}}^2$, both of which are equal to $R_{g,\mathrm{eq}}^2/3$.
Although $R_{g,+}^2(\dot{\epsilon}_\alpha)$ is a monotonically increasing function of $\mathrm{Wi}_\alpha$ regardless of the flow type, $R_{g,+}^2(\dot{\epsilon})$ and $R_{g,+}^2(\dot{\epsilon}_\mathrm{P})$ become larger than $R_{g,+}^2(\dot{\epsilon}_\mathrm{B})$ for $\mathrm{Wi}_\alpha\gtrsim 1$.
As discussed above, the strain rate in the extensional direction is the same for the three extensional flows at fixed $\mathrm{Wi}_\alpha$, and the uniaxial and planar extensional flows have a single extensional direction while biaxial extensional flow has two, thus leading to $R_{g,+}^2(\dot{\epsilon})\simeq R_{g,+}^2(\dot{\epsilon}_\mathrm{P}) >R_{g,+}^2(\dot{\epsilon}_\mathrm{B})$.
In contrast, $R_{g,-}^2(\dot{\epsilon}_\alpha)$ is a monotonically decreasing function of $\mathrm{Wi}_\alpha$, and $R_{g,-}^2(\dot{\epsilon}_\mathrm{B})<R_{g,-}^2(\dot{\epsilon}_\mathrm{P})<R_{g,-}^2(\dot{\epsilon})$ holds according to the order of magnitude of the strain rate in the compressional direction.
Then, combining these observations of the gyration radii and Eq.~\eqref{eq:eta_Rouse_relation_steady} allows us to explain the flow-kinematics dependence of $\eta_{\alpha}(\dot{\epsilon}_\alpha)$.
For $\mathrm{Wi}_\alpha \lesssim 1$, since $R_{g,+}^2(\dot{\epsilon}_\alpha)$ and $R_{g,-}^2(\dot{\epsilon}_\alpha)$ have comparable contributions due to the modest change in polymer conformation, the relation $\eta_\mathrm{B}(\dot{\epsilon}_\mathrm{B}) >\eta_\mathrm{P}(\dot{\epsilon}_\mathrm{P})>\eta_\mathrm{E}(\dot{\epsilon})$ holds, where the difference mainly arises from the kinematic factor $\kappa_\alpha$ in Eq.~\eqref{eq:eta_Rouse_relation_steady}.
It is worth noting that in biaxial extensional flow, $\kappa_\mathrm{B}=2$ and $R_{g,-}^2(\dot{\epsilon}_\mathrm{B})$ decreases relatively rapidly, leading to a slight decrease in $\eta_{\mathrm{B}}(\dot{\epsilon}_\mathrm{B})$~(i.e., $\eta_{\mathrm{B}}(\dot{\epsilon}_\mathrm{B})/\eta_0<6$) for $0.1\lesssim \mathrm{Wi}_\mathrm{B}\lesssim 1$.
For $\mathrm{Wi}_\alpha \gtrsim 1$, $R_{g,+}^2(\dot{\epsilon}_\alpha)$ becomes dominant due to significant polymer stretching, and $\eta_\mathrm{E}(\dot{\epsilon})\simeq \eta_\mathrm{P}(\dot{\epsilon}_\mathrm{P})>\eta_\mathrm{B}(\dot{\epsilon}_\mathrm{B})$ becomes evident owing to the difference in the stretching degree in the extensional direction.
We therefore conclude that the flow-type dependence of the steady-state extensional viscosity arises from (i) the kinematic constant $\kappa_\alpha$ determined by the velocity gradient tensor and (ii) the flow-induced modulation of the gyration radii $R_{g,+}(\dot{\epsilon}_\alpha)$ and $R_{g,-}(\dot{\epsilon}_\alpha)$.
\section{Conclusion}\label{Conclusion}

In this paper, we examined the response of dilute polymer solutions under uniaxial, planar, and biaxial extensional flows using DPD simulations~(Fig.~\ref{fig:snapshot}). 
Dilute polymer solutions exhibit strain hardening during start-up extensional flow at high extension rates, whereas the quantitative behavior depends on the flow kinematics~(Fig.~\ref{fig:vis_growth}).
To elucidate the flow-kinematics dependence, we related the extensional viscosity growth function $\eta_{\alpha,\mathrm{p}}^{+}(t;\dot{\epsilon}_\alpha)$ to polymer conformation using the Rouse-type model.
We demonstrated that $\eta_{\alpha,\mathrm{p}}^{+}(t;\dot{\epsilon}_\alpha)$ is quantitatively reproduced by the instantaneous gyration radii in the extensional~($R_{g,+}(t;\dot{\epsilon}_\alpha)$) and the compressional~($R_{g,-}(t;\dot{\epsilon}_\alpha)$) directions, their time derivatives, and the flow-kinematics dependent constant $\kappa_\alpha$, which reflects the ratio of compression to extension in the imposed flow field.
Using this relation, we explained differences in the extensional viscosity growth function among the three extensional flows in terms of purely kinematic effects and flow-induced polymer conformational changes~(Fig.~\ref{fig:vis_growth_gyration}).

We also demonstrated the flow-kinematics dependence of the steady-state extensional viscosity $\eta_\alpha(\dot{\epsilon}_\alpha)$~(Fig.~\ref{fig:steady_vis}).
At small Weissenberg numbers~$\mathrm{Wi}_\alpha$, $\eta_\mathrm{E}(\dot{\epsilon})<\eta_\mathrm{P}(\dot{\epsilon}_\mathrm{P})<\eta_\mathrm{B}(\dot{\epsilon}_\mathrm{B})$.
In contrast, as $\mathrm{Wi}_\alpha$ increases, the relation $\eta_\mathrm{E}(\dot{\epsilon})\simeq \eta_\mathrm{P}(\dot{\epsilon}_\mathrm{P})>\eta_\mathrm{B}(\dot{\epsilon}_\mathrm{B})$ becomes gradually evident.
The Rouse-type model also provides a relation between the polymer contribution $\eta_{\alpha,\mathrm{p}}(\dot{\epsilon}_\alpha)$ to $\eta_\alpha(\dot{\epsilon}_\alpha)$ and the gyration radii in the extensional~($R_{g,+}(\dot{\epsilon}_\alpha)$) and the compressional~($R_{g,-}(\dot{\epsilon}_\alpha)$) directions~(Fig.~\ref{fig:steady_vis_gyration}).
On the basis of the relation, we explained the flow-kinematics dependence of $\eta_\alpha(\dot{\epsilon}_\alpha)$ in terms of $\mathrm{Wi}_\alpha$-dependent changes in $R_{g,+}(\dot{\epsilon}_\alpha)$ and $R_{g,-}(\dot{\epsilon}_\alpha)$~(Fig.~\ref{fig:gyration}).

In the present study, we established a framework for analyzing the flow-kinematics dependence of the extensional viscosity by focusing on dilute polymer solutions, which serve as a relatively simple starting point.
An important future direction would be to apply our approach to more complex systems such as wormlike micellar solutions~\cite{Koide2025-zr}, which may also exhibit flow-kinematics-dependent scission and recombination kinetics.
\backmatter


\bmhead{Acknowledgements}
This work was supported by JSPS Grants-in-Aid for Scientific Research (24KJ0109) and JST ACT-X (JPMJAX24D5). 
The DPD simulations were mainly conducted under the HPCI Research Projects using the supercomputer ``Flow'' at Information Technology Center, Nagoya University.





\bigskip


\begin{thebibliography}{58}
\ifx \bisbn   \undefined \def \bisbn  #1{ISBN #1}\fi
\ifx \binits  \undefined \def \binits#1{#1}\fi
\ifx \bauthor  \undefined \def \bauthor#1{#1}\fi
\ifx \batitle  \undefined \def \batitle#1{#1}\fi
\ifx \bjtitle  \undefined \def \bjtitle#1{#1}\fi
\ifx \bvolume  \undefined \def \bvolume#1{\textbf{#1}}\fi
\ifx \byear  \undefined \def \byear#1{#1}\fi
\ifx \bissue  \undefined \def \bissue#1{#1}\fi
\ifx \bfpage  \undefined \def \bfpage#1{#1}\fi
\ifx \blpage  \undefined \def \blpage #1{#1}\fi
\ifx \burl  \undefined \def \burl#1{\textsf{#1}}\fi
\ifx \doiurl  \undefined \def \doiurl#1{\url{https://doi.org/#1}}\fi
\ifx \betal  \undefined \def \betal{\textit{et al.}}\fi
\ifx \binstitute  \undefined \def \binstitute#1{#1}\fi
\ifx \binstitutionaled  \undefined \def \binstitutionaled#1{#1}\fi
\ifx \bctitle  \undefined \def \bctitle#1{#1}\fi
\ifx \beditor  \undefined \def \beditor#1{#1}\fi
\ifx \bpublisher  \undefined \def \bpublisher#1{#1}\fi
\ifx \bbtitle  \undefined \def \bbtitle#1{#1}\fi
\ifx \bedition  \undefined \def \bedition#1{#1}\fi
\ifx \bseriesno  \undefined \def \bseriesno#1{#1}\fi
\ifx \blocation  \undefined \def \blocation#1{#1}\fi
\ifx \bsertitle  \undefined \def \bsertitle#1{#1}\fi
\ifx \bsnm \undefined \def \bsnm#1{#1}\fi
\ifx \bsuffix \undefined \def \bsuffix#1{#1}\fi
\ifx \bparticle \undefined \def \bparticle#1{#1}\fi
\ifx \barticle \undefined \def \barticle#1{#1}\fi
\bibcommenthead
\ifx \bconfdate \undefined \def \bconfdate #1{#1}\fi
\ifx \botherref \undefined \def \botherref #1{#1}\fi
\ifx \url \undefined \def \url#1{\textsf{#1}}\fi
\ifx \bchapter \undefined \def \bchapter#1{#1}\fi
\ifx \bbook \undefined \def \bbook#1{#1}\fi
\ifx \bcomment \undefined \def \bcomment#1{#1}\fi
\ifx \oauthor \undefined \def \oauthor#1{#1}\fi
\ifx \citeauthoryear \undefined \def \citeauthoryear#1{#1}\fi
\ifx \endbibitem  \undefined \def \endbibitem {}\fi
\ifx \bconflocation  \undefined \def \bconflocation#1{#1}\fi
\ifx \arxivurl  \undefined \def \arxivurl#1{\textsf{#1}}\fi
\csname PreBibitemsHook\endcsname

\bibitem[\protect\citeauthoryear{Toms}{1949}]{Toms}
\begin{barticle}
\bauthor{\bsnm{Toms}, \binits{B.A.}}:
\batitle{Some observations on the flow of linear polymer solutions through
  straight tubes at large {R}eynolds numbers}.
\bjtitle{in \textit{Proceedings of the 1st International Congress on Rheology}}
\bvolume{2},
\bfpage{135}--\blpage{141}
(\byear{1949})
\end{barticle}
\endbibitem

\bibitem[\protect\citeauthoryear{White and Mungal}{2008}]{White2008-gb}
\begin{barticle}
\bauthor{\bsnm{White}, \binits{C.M.}},
\bauthor{\bsnm{Mungal}, \binits{M.G.}}:
\batitle{Mechanics and prediction of turbulent drag reduction with polymer
  additives}.
\bjtitle{Annu. Rev. Fluid Mech.}
\bvolume{40},
\bfpage{235}--\blpage{256}
(\byear{2008})
\end{barticle}
\endbibitem

\bibitem[\protect\citeauthoryear{Xi}{2019}]{Xi2019-nc}
\begin{barticle}
\bauthor{\bsnm{Xi}, \binits{L.}}:
\batitle{Turbulent drag reduction by polymer additives: Fundamentals and recent
  advances}.
\bjtitle{Phys. Fluids}
\bvolume{31}(\bissue{12}),
\bfpage{121302}
(\byear{2019})
\end{barticle}
\endbibitem

\bibitem[\protect\citeauthoryear{Datta et~al.}{2022}]{Datta2022-vl}
\begin{barticle}
\bauthor{\bsnm{Datta}, \binits{S.S.}},
\bauthor{\bsnm{Ardekani}, \binits{A.M.}},
\bauthor{\bsnm{Arratia}, \binits{P.E.}},
\bauthor{\bsnm{Beris}, \binits{A.N.}},
\bauthor{\bsnm{Bischofberger}, \binits{I.}},
\bauthor{\bsnm{McKinley}, \binits{G.H.}},
\bauthor{\bsnm{Eggers}, \binits{J.G.}},
\bauthor{\bsnm{López-Aguilar}, \binits{J.E.}},
\bauthor{\bsnm{Fielding}, \binits{S.M.}},
\bauthor{\bsnm{Frishman}, \binits{A.}},
\bauthor{\bsnm{Graham}, \binits{M.D.}},
\bauthor{\bsnm{Guasto}, \binits{J.S.}},
\bauthor{\bsnm{Haward}, \binits{S.J.}},
\bauthor{\bsnm{Shen}, \binits{A.Q.}},
\bauthor{\bsnm{Hormozi}, \binits{S.}},
\bauthor{\bsnm{Morozov}, \binits{A.}},
\bauthor{\bsnm{Poole}, \binits{R.J.}},
\bauthor{\bsnm{Shankar}, \binits{V.}},
\bauthor{\bsnm{Shaqfeh}, \binits{E.S.G.}},
\bauthor{\bsnm{Stark}, \binits{H.}},
\bauthor{\bsnm{Steinberg}, \binits{V.}},
\bauthor{\bsnm{Subramanian}, \binits{G.}},
\bauthor{\bsnm{Stone}, \binits{H.A.}}:
\batitle{Perspectives on viscoelastic flow instabilities and elastic
  turbulence}.
\bjtitle{Phys. Rev. Fluids}
\bvolume{7}(\bissue{8}),
\bfpage{080701}
(\byear{2022})
\end{barticle}
\endbibitem

\bibitem[\protect\citeauthoryear{Sasmal}{2025}]{Sasmal2025-pb}
\begin{barticle}
\bauthor{\bsnm{Sasmal}, \binits{C.}}:
\batitle{Potential applications of elastic instability and elastic turbulence:
  A comprehensive review, limitations, and future directions}.
\bjtitle{J. Non-Newtonian Fluid Mech.}
\bvolume{337},
\bfpage{105393}
(\byear{2025})
\end{barticle}
\endbibitem

\bibitem[\protect\citeauthoryear{Perkins et~al.}{1997}]{Perkins1997-dm}
\begin{barticle}
\bauthor{\bsnm{Perkins}, \binits{T.T.}},
\bauthor{\bsnm{Smith}, \binits{D.E.}},
\bauthor{\bsnm{Chu}, \binits{S.}}:
\batitle{Single polymer dynamics in an elongational flow}.
\bjtitle{Science}
\bvolume{276}(\bissue{5321}),
\bfpage{2016}--\blpage{2021}
(\byear{1997})
\end{barticle}
\endbibitem

\bibitem[\protect\citeauthoryear{Smith et~al.}{1999}]{Smith1999-rk}
\begin{barticle}
\bauthor{\bsnm{Smith}, \binits{D.E.}},
\bauthor{\bsnm{Babcock}, \binits{H.P.}},
\bauthor{\bsnm{Chu}, \binits{S.}}:
\batitle{Single-polymer dynamics in steady shear flow}.
\bjtitle{Science}
\bvolume{283}(\bissue{5408}),
\bfpage{1724}--\blpage{1727}
(\byear{1999})
\end{barticle}
\endbibitem

\bibitem[\protect\citeauthoryear{Jones et~al.}{1987}]{Jones1987-gx}
\begin{barticle}
\bauthor{\bsnm{Jones}, \binits{D.M.}},
\bauthor{\bsnm{Walters}, \binits{K.}},
\bauthor{\bsnm{Williams}, \binits{P.R.}}:
\batitle{On the extensional viscosity of mobile polymer solutions}.
\bjtitle{Rheol. Acta}
\bvolume{26},
\bfpage{20}--\blpage{30}
(\byear{1987})
\end{barticle}
\endbibitem

\bibitem[\protect\citeauthoryear{Haward et~al.}{2023}]{Haward2023b}
\begin{barticle}
\bauthor{\bsnm{Haward}, \binits{S.J.}},
\bauthor{\bsnm{Varchanis}, \binits{S.}},
\bauthor{\bsnm{McKinley}, \binits{G.H.}},
\bauthor{\bsnm{Alves}, \binits{M.A.}},
\bauthor{\bsnm{Shen}, \binits{A.Q.}}:
\batitle{{Extensional rheometry of mobile fluids. Part II: Comparison between
  the uniaxial, planar, and biaxial extensional rheology of dilute polymer
  solutions using numerically optimized stagnation point microfluidic
  devices}}.
\bjtitle{J. Rheol.}
\bvolume{67}(\bissue{5}),
\bfpage{1011}--\blpage{1030}
(\byear{2023})
\end{barticle}
\endbibitem

\bibitem[\protect\citeauthoryear{Khan et~al.}{1987}]{Khan1987-br}
\begin{barticle}
\bauthor{\bsnm{Khan}, \binits{S.A.}},
\bauthor{\bsnm{Prud'homme}, \binits{R.K.}},
\bauthor{\bsnm{Larson}, \binits{R.G.}}:
\batitle{Comparison of the rheology of polymer melts in shear, and biaxial and
  uniaxial extensions}.
\bjtitle{Rheol. Acta}
\bvolume{26},
\bfpage{144}--\blpage{151}
(\byear{1987})
\end{barticle}
\endbibitem

\bibitem[\protect\citeauthoryear{Takahashi et~al.}{1993}]{Takahashi1993-mb}
\begin{barticle}
\bauthor{\bsnm{Takahashi}, \binits{M.}},
\bauthor{\bsnm{Isaki}, \binits{T.}},
\bauthor{\bsnm{Takigawa}, \binits{T.}},
\bauthor{\bsnm{Masuda}, \binits{T.}}:
\batitle{Measurement of biaxial and uniaxial extensional flow behavior of
  polymer melts at constant strain rates}.
\bjtitle{J. Rheol.}
\bvolume{37}(\bissue{5}),
\bfpage{827}--\blpage{846}
(\byear{1993})
\end{barticle}
\endbibitem

\bibitem[\protect\citeauthoryear{Nishioka et~al.}{2000}]{Nishioka2000-eh}
\begin{barticle}
\bauthor{\bsnm{Nishioka}, \binits{A.}},
\bauthor{\bsnm{Takahashi}, \binits{T.}},
\bauthor{\bsnm{Masubuchi}, \binits{Y.}},
\bauthor{\bsnm{Takimoto}, \binits{J.}},
\bauthor{\bsnm{Koyama}, \binits{K.}}:
\batitle{Description of uniaxial, biaxial, and planar elongational viscosities
  of polystyrene melt by the {K}-{BKZ} model}.
\bjtitle{J. Non-Newtonian Fluid Mech.}
\bvolume{89}(\bissue{3}),
\bfpage{287}--\blpage{301}
(\byear{2000})
\end{barticle}
\endbibitem

\bibitem[\protect\citeauthoryear{Wagner et~al.}{2001}]{Wagner2001-by}
\begin{barticle}
\bauthor{\bsnm{Wagner}, \binits{M.H.}},
\bauthor{\bsnm{Rubio}, \binits{P.}},
\bauthor{\bsnm{Bastian}, \binits{H.}}:
\batitle{The molecular stress function model for polydisperse polymer melts
  with dissipative convective constraint release}.
\bjtitle{J. Rheol.}
\bvolume{45}(\bissue{6}),
\bfpage{1387}--\blpage{1412}
(\byear{2001})
\end{barticle}
\endbibitem

\bibitem[\protect\citeauthoryear{Sugimoto et~al.}{2001}]{Sugimoto2001-ji}
\begin{barticle}
\bauthor{\bsnm{Sugimoto}, \binits{M.}},
\bauthor{\bsnm{Masubuchi}, \binits{Y.}},
\bauthor{\bsnm{Takimoto}, \binits{J.}},
\bauthor{\bsnm{Koyama}, \binits{K.}}:
\batitle{Melt rheology of polypropylene containing small amounts of
  high-molecular-weight chain. 2. uniaxial and biaxial extensional flow}.
\bjtitle{Macromolecules}
\bvolume{34}(\bissue{17}),
\bfpage{6056}--\blpage{6063}
(\byear{2001})
\end{barticle}
\endbibitem

\bibitem[\protect\citeauthoryear{Hachmann and Meissner}{2003}]{Hachmann2003-di}
\begin{barticle}
\bauthor{\bsnm{Hachmann}, \binits{P.}},
\bauthor{\bsnm{Meissner}, \binits{J.}}:
\batitle{Rheometer for equibiaxial and planar elongations of polymer melts}.
\bjtitle{J. Rheol.}
\bvolume{47}(\bissue{4}),
\bfpage{989}--\blpage{1010}
(\byear{2003})
\end{barticle}
\endbibitem

\bibitem[\protect\citeauthoryear{Stadler et~al.}{2007}]{Stadler2007-gq}
\begin{barticle}
\bauthor{\bsnm{Stadler}, \binits{F.J.}},
\bauthor{\bsnm{Nishioka}, \binits{A.}},
\bauthor{\bsnm{Stange}, \binits{J.}},
\bauthor{\bsnm{Koyama}, \binits{K.}},
\bauthor{\bsnm{Münstedt}, \binits{H.}}:
\batitle{Comparison of the elongational behavior of various polyolefins in
  uniaxial and equibiaxial flows}.
\bjtitle{Rheol. Acta}
\bvolume{46},
\bfpage{1003}--\blpage{1012}
(\byear{2007})
\end{barticle}
\endbibitem

\bibitem[\protect\citeauthoryear{McKinley and Tripathi}{2000}]{McKinley2000-ky}
\begin{barticle}
\bauthor{\bsnm{McKinley}, \binits{G.H.}},
\bauthor{\bsnm{Tripathi}, \binits{A.}}:
\batitle{How to extract the {N}ewtonian viscosity from capillary breakup
  measurements in a filament rheometer}.
\bjtitle{J. Rheol.}
\bvolume{44}(\bissue{3}),
\bfpage{653}--\blpage{670}
(\byear{2000})
\end{barticle}
\endbibitem

\bibitem[\protect\citeauthoryear{Anna and McKinley}{2001}]{Anna2001-bn}
\begin{barticle}
\bauthor{\bsnm{Anna}, \binits{S.L.}},
\bauthor{\bsnm{McKinley}, \binits{G.H.}}:
\batitle{Elasto-capillary thinning and breakup of model elastic liquids}.
\bjtitle{J. Rheol.}
\bvolume{45}(\bissue{1}),
\bfpage{115}--\blpage{138}
(\byear{2001})
\end{barticle}
\endbibitem

\bibitem[\protect\citeauthoryear{Haward}{2016}]{Haward2016-lh}
\begin{barticle}
\bauthor{\bsnm{Haward}, \binits{S.J.}}:
\batitle{Microfluidic extensional rheometry using stagnation point flow}.
\bjtitle{Biomicrofluidics}
\bvolume{10}(\bissue{4}),
\bfpage{043401}
(\byear{2016})
\end{barticle}
\endbibitem

\bibitem[\protect\citeauthoryear{Haward et~al.}{2023}]{Haward2023a}
\begin{barticle}
\bauthor{\bsnm{Haward}, \binits{S.J.}},
\bauthor{\bsnm{Pimenta}, \binits{F.}},
\bauthor{\bsnm{Varchanis}, \binits{S.}},
\bauthor{\bsnm{Carlson}, \binits{D.W.}},
\bauthor{\bsnm{Toda-Peters}, \binits{K.}},
\bauthor{\bsnm{Alves}, \binits{M.A.}},
\bauthor{\bsnm{Shen}, \binits{A.Q.}}:
\batitle{{Extensional rheometry of mobile fluids. Part I: OUBER, an optimized
  uniaxial and biaxial extensional rheometer}}.
\bjtitle{J. Rheol.}
\bvolume{67}(\bissue{5}),
\bfpage{995}--\blpage{1009}
(\byear{2023})
\end{barticle}
\endbibitem

\bibitem[\protect\citeauthoryear{van~den Brule}{1993}]{van-den-Brule1993-mg}
\begin{barticle}
\bauthor{\bsnm{Brule}, \binits{B.H.A.A.}}:
\batitle{Brownian dynamics simulation of finitely extensible bead-spring
  chains}.
\bjtitle{J. Non-Newtonian Fluid Mech.}
\bvolume{47},
\bfpage{357}--\blpage{378}
(\byear{1993})
\end{barticle}
\endbibitem

\bibitem[\protect\citeauthoryear{Doyle et~al.}{1997}]{Doyle1997-tq}
\begin{barticle}
\bauthor{\bsnm{Doyle}, \binits{P.S.}},
\bauthor{\bsnm{Shaqfeh}, \binits{E.S.G.}},
\bauthor{\bsnm{Gast}, \binits{A.P.}}:
\batitle{Dynamic simulation of freely draining flexible polymers in steady
  linear flows}.
\bjtitle{J. Fluid Mech.}
\bvolume{334},
\bfpage{251}--\blpage{291}
(\byear{1997})
\end{barticle}
\endbibitem

\bibitem[\protect\citeauthoryear{Herrchen and
  Öttinger}{1997}]{Herrchen1997-fb}
\begin{barticle}
\bauthor{\bsnm{Herrchen}, \binits{M.}},
\bauthor{\bsnm{Öttinger}, \binits{H.C.}}:
\batitle{A detailed comparison of various {FENE} dumbbell models}.
\bjtitle{J. Non-Newtonian Fluid Mech.}
\bvolume{68}(\bissue{1}),
\bfpage{17}--\blpage{42}
(\byear{1997})
\end{barticle}
\endbibitem

\bibitem[\protect\citeauthoryear{Li et~al.}{2000}]{Li2000-da}
\begin{barticle}
\bauthor{\bsnm{Li}, \binits{L.}},
\bauthor{\bsnm{Larson}, \binits{R.G.}},
\bauthor{\bsnm{Sridhar}, \binits{T.}}:
\batitle{Brownian dynamics simulations of dilute polystyrene solutions}.
\bjtitle{J. Rheol.}
\bvolume{44}(\bissue{2}),
\bfpage{291}--\blpage{322}
(\byear{2000})
\end{barticle}
\endbibitem

\bibitem[\protect\citeauthoryear{Larson et~al.}{1999}]{Larson2002-mw}
\begin{barticle}
\bauthor{\bsnm{Larson}, \binits{R.G.}},
\bauthor{\bsnm{Hu}, \binits{H.}},
\bauthor{\bsnm{Smith}, \binits{D.E.}},
\bauthor{\bsnm{Chu}, \binits{S.}}:
\batitle{Brownian dynamics simulations of a {DNA} molecule in an extensional
  flow field}.
\bjtitle{J. Rheol.}
\bvolume{43}(\bissue{2}),
\bfpage{267}--\blpage{304}
(\byear{1999})
\end{barticle}
\endbibitem

\bibitem[\protect\citeauthoryear{Somasi et~al.}{2002}]{Somasi2002-ot}
\begin{barticle}
\bauthor{\bsnm{Somasi}, \binits{M.}},
\bauthor{\bsnm{Khomami}, \binits{B.}},
\bauthor{\bsnm{Woo}, \binits{N.J.}},
\bauthor{\bsnm{Hur}, \binits{J.S.}},
\bauthor{\bsnm{Shaqfeh}, \binits{E.S.G.}}:
\batitle{Brownian dynamics simulations of bead-rod and bead-spring chains:
  Numerical algorithms and coarse-graining issues}.
\bjtitle{J. Non-Newtonian Fluid Mech.}
\bvolume{108}(\bissue{1-3}),
\bfpage{227}--\blpage{255}
(\byear{2002})
\end{barticle}
\endbibitem

\bibitem[\protect\citeauthoryear{Hsieh et~al.}{2003}]{Hsieh2003-oc}
\begin{barticle}
\bauthor{\bsnm{Hsieh}, \binits{C.-C.}},
\bauthor{\bsnm{Li}, \binits{L.}},
\bauthor{\bsnm{Larson}, \binits{R.G.}}:
\batitle{Modeling hydrodynamic interaction in {B}rownian dynamics: simulations
  of extensional flows of dilute solutions of {DNA} and polystyrene}.
\bjtitle{J. Non-Newtonian Fluid Mech.}
\bvolume{113}(\bissue{2-3}),
\bfpage{147}--\blpage{191}
(\byear{2003})
\end{barticle}
\endbibitem

\bibitem[\protect\citeauthoryear{Larson}{2005}]{Larson2005-ef}
\begin{barticle}
\bauthor{\bsnm{Larson}, \binits{R.G.}}:
\batitle{The rheology of dilute solutions of flexible polymers: Progress and
  problems}.
\bjtitle{J. Rheol.}
\bvolume{49}(\bissue{1}),
\bfpage{1}--\blpage{70}
(\byear{2005})
\end{barticle}
\endbibitem

\bibitem[\protect\citeauthoryear{Koide et~al.}{2025}]{koide2025relation}
\begin{botherref}
\oauthor{\bsnm{Koide}, \binits{Y.}},
\oauthor{\bsnm{Ishida}, \binits{T.}},
\oauthor{\bsnm{Uneyama}, \binits{T.}},
\oauthor{\bsnm{Masubuchi}, \binits{Y.}}:
Relation between extensional viscosity and polymer conformation in dilute
  polymer solutions.
arXiv:2511.21077
(2025)
\end{botherref}
\endbibitem

\bibitem[\protect\citeauthoryear{Takeda et~al.}{2015}]{Takeda2015-ub}
\begin{barticle}
\bauthor{\bsnm{Takeda}, \binits{K.}},
\bauthor{\bsnm{Sukumaran}, \binits{S.K.}},
\bauthor{\bsnm{Sugimoto}, \binits{M.}},
\bauthor{\bsnm{Koyama}, \binits{K.}},
\bauthor{\bsnm{Masubuchi}, \binits{Y.}}:
\batitle{Test of the stretch/orientation-induced reduction of friction for
  biaxial elongational flow via primitive chain network simulation}.
\bjtitle{Nihon Reoroji Gakkaishi}
\bvolume{43}(\bissue{3-4}),
\bfpage{63}--\blpage{69}
(\byear{2015})
\end{barticle}
\endbibitem

\bibitem[\protect\citeauthoryear{Murashima et~al.}{2018}]{Murashima2018-te}
\begin{barticle}
\bauthor{\bsnm{Murashima}, \binits{T.}},
\bauthor{\bsnm{Hagita}, \binits{K.}},
\bauthor{\bsnm{Kawakatsu}, \binits{T.}}:
\batitle{Elongational viscosity of weakly entangled polymer melt via
  coarse-grained molecular dynamics simulation}.
\bjtitle{Nihon Reoroji Gakkaishi}
\bvolume{46}(\bissue{5}),
\bfpage{207}--\blpage{220}
(\byear{2018})
\end{barticle}
\endbibitem

\bibitem[\protect\citeauthoryear{Murashima et~al.}{2021}]{Murashima2021-wh}
\begin{barticle}
\bauthor{\bsnm{Murashima}, \binits{T.}},
\bauthor{\bsnm{Hagita}, \binits{K.}},
\bauthor{\bsnm{Kawakatsu}, \binits{T.}}:
\batitle{Viscosity overshoot in biaxial elongational flow: Coarse-grained
  molecular dynamics simulation of ring--linear polymer mixtures}.
\bjtitle{Macromolecules}
\bvolume{54}(\bissue{15}),
\bfpage{7210}--\blpage{7225}
(\byear{2021})
\end{barticle}
\endbibitem

\bibitem[\protect\citeauthoryear{Zhou and Akhavan}{2003}]{Zhou2003-gk}
\begin{barticle}
\bauthor{\bsnm{Zhou}, \binits{Q.}},
\bauthor{\bsnm{Akhavan}, \binits{R.}}:
\batitle{A comparison of {FENE} and {FENE-P} dumbbell and chain models in
  turbulent flow}.
\bjtitle{J. Non-Newtonian Fluid Mech.}
\bvolume{109}(\bissue{2-3}),
\bfpage{115}--\blpage{155}
(\byear{2003})
\end{barticle}
\endbibitem

\bibitem[\protect\citeauthoryear{Terrapon et~al.}{2004}]{Terrapon2004-xb}
\begin{barticle}
\bauthor{\bsnm{Terrapon}, \binits{V.E.}},
\bauthor{\bsnm{Dubief}, \binits{Y.}},
\bauthor{\bsnm{Moin}, \binits{P.}},
\bauthor{\bsnm{Shaqfeh}, \binits{E.S.G.}},
\bauthor{\bsnm{Lele}, \binits{S.K.}}:
\batitle{Simulated polymer stretch in a turbulent flow using {B}rownian
  dynamics}.
\bjtitle{J. Fluid Mech.}
\bvolume{504},
\bfpage{61}--\blpage{71}
(\byear{2004})
\end{barticle}
\endbibitem

\bibitem[\protect\citeauthoryear{Jiang et~al.}{2007}]{Jiang2007-rf}
\begin{barticle}
\bauthor{\bsnm{Jiang}, \binits{W.}},
\bauthor{\bsnm{Huang}, \binits{J.}},
\bauthor{\bsnm{Wang}, \binits{Y.}},
\bauthor{\bsnm{Laradji}, \binits{M.}}:
\batitle{Hydrodynamic interaction in polymer solutions simulated with
  dissipative particle dynamics}.
\bjtitle{J. Chem. Phys.}
\bvolume{126}(\bissue{4}),
\bfpage{044901}
(\byear{2007})
\end{barticle}
\endbibitem

\bibitem[\protect\citeauthoryear{Koide and Goto}{2022}]{Koide2022-bp}
\begin{barticle}
\bauthor{\bsnm{Koide}, \binits{Y.}},
\bauthor{\bsnm{Goto}, \binits{S.}}:
\batitle{Flow-induced scission of wormlike micelles in nonionic surfactant
  solutions under shear flow}.
\bjtitle{J. Chem. Phys.}
\bvolume{157}(\bissue{8}),
\bfpage{084903}
(\byear{2022})
\end{barticle}
\endbibitem

\bibitem[\protect\citeauthoryear{Koide and Goto}{2023}]{Koide2023-ao}
\begin{barticle}
\bauthor{\bsnm{Koide}, \binits{Y.}},
\bauthor{\bsnm{Goto}, \binits{S.}}:
\batitle{Effect of scission on alignment of nonionic surfactant micelles under
  shear flow}.
\bjtitle{Soft Matter}
\bvolume{19}(\bissue{23}),
\bfpage{4323}--\blpage{4332}
(\byear{2023})
\end{barticle}
\endbibitem

\bibitem[\protect\citeauthoryear{Koide et~al.}{2025}]{Koide2025-zr}
\begin{barticle}
\bauthor{\bsnm{Koide}, \binits{Y.}},
\bauthor{\bsnm{Ishida}, \binits{T.}},
\bauthor{\bsnm{Uneyama}, \binits{T.}},
\bauthor{\bsnm{Masubuchi}, \binits{Y.}}:
\batitle{Steady-state extensional viscosity of wormlike micellar solutions via
  dissipative particle dynamics simulations}.
\bjtitle{Macromolecules}
\bvolume{58}(\bissue{22}),
\bfpage{12039}--\blpage{12051}
(\byear{2025})
\end{barticle}
\endbibitem

\bibitem[\protect\citeauthoryear{Evans and Morriss}{2008}]{evans_morriss_2008}
\begin{bbook}
\bauthor{\bsnm{Evans}, \binits{D.J.}},
\bauthor{\bsnm{Morriss}, \binits{G.P.}}:
\bbtitle{Statistical Mechanics of Nonequilibrium Liquids}.
\bpublisher{Cambridge University Press},
\blocation{Cambridge}
(\byear{2008})
\end{bbook}
\endbibitem

\bibitem[\protect\citeauthoryear{Dobson}{2014}]{Dobson2014-kr}
\begin{barticle}
\bauthor{\bsnm{Dobson}, \binits{M.}}:
\batitle{Periodic boundary conditions for long-time nonequilibrium molecular
  dynamics simulations of incompressible flows}.
\bjtitle{J. Chem. Phys.}
\bvolume{141}(\bissue{18}),
\bfpage{184103}
(\byear{2014})
\end{barticle}
\endbibitem

\bibitem[\protect\citeauthoryear{Hunt}{2016}]{Hunt2016-vl}
\begin{barticle}
\bauthor{\bsnm{Hunt}, \binits{T.A.}}:
\batitle{Periodic boundary conditions for the simulation of uniaxial
  extensional flow of arbitrary duration}.
\bjtitle{Mol. Simul.}
\bvolume{42}(\bissue{5}),
\bfpage{347}--\blpage{352}
(\byear{2016})
\end{barticle}
\endbibitem

\bibitem[\protect\citeauthoryear{Semaev}{2001}]{Semaev2001-sn}
\begin{bchapter}
\bauthor{\bsnm{Semaev}, \binits{I.}}:
\bctitle{A 3-dimensional lattice reduction algorithm}.
In: \beditor{\bsnm{Silverman}, \binits{J.H.}} (ed.)
\bbtitle{Cryptography and Lattices},
pp. \bfpage{181}--\blpage{193}.
\bpublisher{Springer},
\blocation{Berlin, Heidelberg}
(\byear{2001})
\end{bchapter}
\endbibitem

\bibitem[\protect\citeauthoryear{Todd and Daivis}{2000}]{Todd2000-rh}
\begin{barticle}
\bauthor{\bsnm{Todd}, \binits{B.D.}},
\bauthor{\bsnm{Daivis}, \binits{P.J.}}:
\batitle{The stability of nonequilibrium molecular dynamics simulations of
  elongational flows}.
\bjtitle{J. Chem. Phys.}
\bvolume{112}(\bissue{1}),
\bfpage{40}--\blpage{46}
(\byear{2000})
\end{barticle}
\endbibitem

\bibitem[\protect\citeauthoryear{Nicholson and
  Rutledge}{2016}]{Nicholson2016-aj}
\begin{barticle}
\bauthor{\bsnm{Nicholson}, \binits{D.A.}},
\bauthor{\bsnm{Rutledge}, \binits{G.C.}}:
\batitle{Molecular simulation of flow-enhanced nucleation in
  \textit{n}-eicosane melts under steady shear and uniaxial extension}.
\bjtitle{J. Chem. Phys.}
\bvolume{145}(\bissue{24}),
\bfpage{244903}
(\byear{2016})
\end{barticle}
\endbibitem

\bibitem[\protect\citeauthoryear{Groot and Warren}{1997}]{Groot1997-je}
\begin{barticle}
\bauthor{\bsnm{Groot}, \binits{R.D.}},
\bauthor{\bsnm{Warren}, \binits{P.B.}}:
\batitle{Dissipative particle dynamics: Bridging the gap between atomistic and
  mesoscopic simulation}.
\bjtitle{J. Chem. Phys.}
\bvolume{107}(\bissue{11}),
\bfpage{4423}--\blpage{4435}
(\byear{1997})
\end{barticle}
\endbibitem

\bibitem[\protect\citeauthoryear{Irving and Kirkwood}{1950}]{Irving1950-lc}
\begin{barticle}
\bauthor{\bsnm{Irving}, \binits{J.H.}},
\bauthor{\bsnm{Kirkwood}, \binits{J.G.}}:
\batitle{The statistical mechanical theory of transport processes. {IV}. {T}he
  equations of hydrodynamics}.
\bjtitle{J. Chem. Phys.}
\bvolume{18}(\bissue{6}),
\bfpage{817}--\blpage{829}
(\byear{1950})
\end{barticle}
\endbibitem

\bibitem[\protect\citeauthoryear{Liu et~al.}{2015}]{Liu2015-yj}
\begin{barticle}
\bauthor{\bsnm{Liu}, \binits{M.B.}},
\bauthor{\bsnm{Liu}, \binits{G.R.}},
\bauthor{\bsnm{Zhou}, \binits{L.W.}},
\bauthor{\bsnm{Chang}, \binits{J.Z.}}:
\batitle{Dissipative particle dynamics ({DPD}): An overview and recent
  developments}.
\bjtitle{Arch. Comput. Methods Eng.}
\bvolume{22},
\bfpage{529}--\blpage{556}
(\byear{2015})
\end{barticle}
\endbibitem

\bibitem[\protect\citeauthoryear{Sridhar et~al.}{1991}]{Sridhar1991-th}
\begin{barticle}
\bauthor{\bsnm{Sridhar}, \binits{T.}},
\bauthor{\bsnm{Tirtaatmadja}, \binits{V.}},
\bauthor{\bsnm{Nguyen}, \binits{D.A.}},
\bauthor{\bsnm{Gupta}, \binits{R.K.}}:
\batitle{Measurement of extensional viscosity of polymer solutions}.
\bjtitle{J. Non-Newtonian Fluid Mech.}
\bvolume{40}(\bissue{3}),
\bfpage{271}--\blpage{280}
(\byear{1991})
\end{barticle}
\endbibitem

\bibitem[\protect\citeauthoryear{Gupta et~al.}{2000}]{Gupta2000-ad}
\begin{barticle}
\bauthor{\bsnm{Gupta}, \binits{R.K.}},
\bauthor{\bsnm{Nguyen}, \binits{D.A.}},
\bauthor{\bsnm{Sridhar}, \binits{T.}}:
\batitle{Extensional viscosity of dilute polystyrene solutions: Effect of
  concentration and molecular weight}.
\bjtitle{Phys. Fluids}
\bvolume{12}(\bissue{6}),
\bfpage{1296}--\blpage{1318}
(\byear{2000})
\end{barticle}
\endbibitem

\bibitem[\protect\citeauthoryear{Anna et~al.}{2001}]{Anna2001-hj}
\begin{barticle}
\bauthor{\bsnm{Anna}, \binits{S.L.}},
\bauthor{\bsnm{McKinley}, \binits{G.H.}},
\bauthor{\bsnm{Nguyen}, \binits{D.A.}},
\bauthor{\bsnm{Sridhar}, \binits{T.}},
\bauthor{\bsnm{Muller}, \binits{S.J.}},
\bauthor{\bsnm{Huang}, \binits{J.}},
\bauthor{\bsnm{James}, \binits{D.F.}}:
\batitle{An interlaboratory comparison of measurements from filament-stretching
  rheometers using common test fluids}.
\bjtitle{J. Rheol.}
\bvolume{45}(\bissue{1}),
\bfpage{83}--\blpage{114}
(\byear{2001})
\end{barticle}
\endbibitem

\bibitem[\protect\citeauthoryear{Uneyama}{2025}]{uneyama2025radius}
\begin{barticle}
\bauthor{\bsnm{Uneyama}, \binits{T.}}:
\batitle{Radius of gyration in shear gradient direction governs steady shear
  viscosity of rouse-type model}.
\bjtitle{Nihon Reoroji Gakkaishi}
\bvolume{53}(\bissue{1}),
\bfpage{11}--\blpage{19}
(\byear{2025})
\end{barticle}
\endbibitem

\bibitem[\protect\citeauthoryear{Rouse}{1953}]{Rouse1953-hp}
\begin{barticle}
\bauthor{\bsnm{Rouse}, \binits{P.E.} \bsuffix{Jr.}}:
\batitle{A theory of the linear viscoelastic properties of dilute solutions of
  coiling polymers}.
\bjtitle{J. Chem. Phys.}
\bvolume{21}(\bissue{7}),
\bfpage{1272}--\blpage{1280}
(\byear{1953})
\end{barticle}
\endbibitem

\bibitem[\protect\citeauthoryear{Bird et~al.}{1987a}]{Bird1987DPLv1}
\begin{bbook}
\bauthor{\bsnm{Bird}, \binits{R.B.}},
\bauthor{\bsnm{Armstrong}, \binits{R.C.}},
\bauthor{\bsnm{Hassager}, \binits{O.}}:
\bbtitle{Dynamics of Polymeric Liquids: Volume 1: Fluid Mechanics},
\bedition{2}nd edn.
\bpublisher{John Wiley \& Sons},
\blocation{New York}
(\byear{1987})
\end{bbook}
\endbibitem

\bibitem[\protect\citeauthoryear{Bird et~al.}{1987b}]{bird1987dynamics}
\begin{bbook}
\bauthor{\bsnm{Bird}, \binits{R.B.}},
\bauthor{\bsnm{Curtiss}, \binits{C.F.}},
\bauthor{\bsnm{Armstrong}, \binits{R.C.}},
\bauthor{\bsnm{Hassager}, \binits{O.}}:
\bbtitle{Dynamics of {P}olymeric {L}iquids, Volume 2: Kinetic Theory},
\bedition{2}nd edn.
\bpublisher{John Wiley \& Sons},
\blocation{New York}
(\byear{1987})
\end{bbook}
\endbibitem

\bibitem[\protect\citeauthoryear{Rubinstein and
  Colby}{2003}]{rubinstein2003polymer}
\begin{bbook}
\bauthor{\bsnm{Rubinstein}, \binits{M.}},
\bauthor{\bsnm{Colby}, \binits{R.H.}}:
\bbtitle{Polymer Physics}.
\bpublisher{Oxford University Press},
\blocation{Oxford}
(\byear{2003})
\end{bbook}
\endbibitem

\bibitem[\protect\citeauthoryear{Prabhakar et~al.}{2016}]{Prabhakar2016-ba}
\begin{barticle}
\bauthor{\bsnm{Prabhakar}, \binits{R.}},
\bauthor{\bsnm{Gadkari}, \binits{S.}},
\bauthor{\bsnm{Gopesh}, \binits{T.}},
\bauthor{\bsnm{Shaw}, \binits{M.J.}}:
\batitle{Influence of stretching induced self-concentration and self-dilution
  on coil-stretch hysteresis and capillary thinning of unentangled polymer
  solutions}.
\bjtitle{J. Rheol.}
\bvolume{60}(\bissue{3}),
\bfpage{345}--\blpage{366}
(\byear{2016})
\end{barticle}
\endbibitem

\bibitem[\protect\citeauthoryear{Ianniruberto
  et~al.}{2020}]{Ianniruberto2020-rg}
\begin{barticle}
\bauthor{\bsnm{Ianniruberto}, \binits{G.}},
\bauthor{\bsnm{Marrucci}, \binits{G.}},
\bauthor{\bsnm{Masubuchi}, \binits{Y.}}:
\batitle{Melts of linear polymers in fast flows}.
\bjtitle{Macromolecules}
\bvolume{53}(\bissue{13}),
\bfpage{5023}--\blpage{5033}
(\byear{2020})
\end{barticle}
\endbibitem

\bibitem[\protect\citeauthoryear{Matsumiya and
  Watanabe}{2021}]{Matsumiya2021-ti}
\begin{barticle}
\bauthor{\bsnm{Matsumiya}, \binits{Y.}},
\bauthor{\bsnm{Watanabe}, \binits{H.}}:
\batitle{Non-universal features in uniaxially extensional rheology of linear
  polymer melts and concentrated solutions: A review}.
\bjtitle{Prog. Polym. Sci.}
\bvolume{112},
\bfpage{101325}
(\byear{2021})
\end{barticle}
\endbibitem

\end{thebibliography}
\end{document}